\documentclass[12pt]{article}
\usepackage{url}
\usepackage{hyperref}
\usepackage{changebar}
\usepackage[small,bf]{caption}
\usepackage{lineno}
\usepackage{subfigure}
\usepackage{epsfig}
\usepackage[top=1.0in, bottom=1.0in, left=1.0in, right=1.0in]{geometry}
\usepackage{mathpple}
\usepackage{authblk}
\newcommand{\pp}{$pp$}
\newcommand{\ppb}{$p$+Pb}
\newcommand{\dau}{$d$+Au}
\newcommand{\snn}{\sqrt{s_{_{NN}}}}
\newcommand{\gev}{GeV/$c$}
\newcommand{\pt}{p_{T}}
\newcommand{\ptref}{p_{T}^{\rm ref}}
\newcommand{\etaref}{\eta^{\rm ref}}
\newcommand{\ptrefa}{p_{T}^{\rm ref_1}}
\newcommand{\etarefa}{\eta^{\rm ref_1}}
\newcommand{\ptrefb}{p_{T}^{\rm ref_2}}
\newcommand{\etarefb}{\eta^{\rm ref_2}}
\newcommand{\trig}{^{\rm trig}}
\newcommand{\asso}{^{\rm assoc}}
\newcommand{\ptt}{p_{T}^{\trig}}
\newcommand{\pta}{p_{T}^{\asso}}
\newcommand{\etat}{\eta^{\trig}}
\newcommand{\etaa}{\eta^{\asso}}
\newcommand{\dphi}{\Delta\phi}
\newcommand{\deta}{\Delta\eta}
\newcommand{\Vn}[1]{V_n\{#1\}}

\author{Fuqiang Wang}
\affil{Department of Physics, Purdue University, West Lafayette, Indiana 47907, USA}
\title{Novel Phenomena in Particle Correlations in Relativistic Heavy-Ion Collisions}
\begin{document}
\date{}
\maketitle
\begin{abstract}
The novel phenomena observed in particle angular correlations are reviewed. They include the double-peak away-side azimuthal correlations in relativistic heavy-ion collisions and the long-range pseudorapidity near-side (ridge) correlations in heavy-ion as well as in proton-induced collisions. The collision system and energy dependence of these phenomena are examined, wherever possible and most abundantly for the ridge correlations. Their possible theoretical interpretations and what might be learned about the properties of the collision systems from theoretical comparisons are discussed. Prospective future measurements and theoretical undertakings are outlined that might help further the understanding of the physics mechanisms underlying these phenomena.
\end{abstract}
\clearpage
\tableofcontents
\section{Introduction}
The primary goal of the nuclear physics programs at Brookhaven National Laboratory's Relativistic Heavy-Ion Collider (RHIC) and CERN's Large Hadron Collider (LHC) is to study high temperature quantum chromodynamics (QCD). Large nuclei are collided at extremely high energies so that the matter created is so hot that nucleons are no longer the relevant degrees of freedom. Instead, quarks and gluons are deconfined over an extended volume, the quark-gluon plasma (QGP)~\cite{Arsene:2004fa,Back:2004je,Adams:2005dq,Adcox:2004mh}. Measurements of the photons radiated from this matter are consistent with its initial temperature in excess of 220~MeV~\cite{Adare:2008ab}, the hottest matter ever created in laboratory. 

In many ways everything about such a system is ``novel.'' The bulk behavior of the matter is well described by hydrodynamic models with a surprisingly small shear viscosity to entropy density ratio, $\eta/s$, not much larger than the conjectured quantum lower limit~\cite{Kovtun:2004de,Romatschke:2007mq,Luzum:2012wu}.  The matter is nearly opaque to the passage of high momentum quarks and gluons~\cite{Jacobs:2004qv}, up to the highest measured momenta~\cite{Chatrchyan:2012nia} and the heaviest measured quarks~\cite{Adare:2010de,Nguyen:2012yx}. High momentum quarks and gluons, produced by hard (large momentum transfer) QCD processes, probe this matter on short length scales and lose energy via gluon bremsstrahlung and multiple scattering~\cite{Gyulassy:2003mc}. 

This review discusses some of the features in particle angular correlations in heavy-ion, proton-proton (\pp), proton-lead (\ppb), and deuteron-gold (\dau) collisions. These correlations have long been used for studies of jet physics.  However, structures very different from di-jet expectations in \pp\ collisions are observed in heavy-ion collisions, largely unexpected prior to the start of RHIC data taking in 2000. In particular, a small azimuthal angle (near-side) large pseudorapidity correlation, the {\it ridge} and a broadening in large azimuthal angle (away-side) correlation are observed. The ridge is also observed in large-multiplicity \pp, \ppb, and \dau\ collisions. These phenomena encompass the forefront of theoretical understanding and calculational ability and require creative and differential experimental techniques to extract from the remnants of the collision. 
This article reviews the experimental measurements of these novel phenomena in particle correlations, their possible theoretical interpretations, and what future measurements and theoretical developments may be needed to further our understanding of those phenomena in connection to the properties of the QGP. 
\section{Angular Correlations Preamble}
High-$\pt$ particle production  has been measured to be strongly suppressed in heavy-ion collisions relative to \pp\ collisions~\cite{Adcox:2001jp,Adler:2002xw,Adler:2003qi,Adams:2003kv}. This jet-quenching phenomenon is understood to be caused by parton energy loss in the high density medium created in heavy-ion collisions~\cite{Gyulassy:1990ye,Wang:1991xy}. Because the parton energy distribution is an steeply falling function of energy, the surviving particles at high $\pt$ are likely those with minimal interactions with the medium.  Since the medium interaction probability increases with the path length through the matter, measurements of single high $p_T$ hadrons are thought to be largely sensitive to partons directed outward from the near-surface region of the collision zone. This effect is often called surface bias. The study of single particle production at high $\pt$, therefore, gives limited information about the interior of the QCD medium and about jet-medium interactions. However, jet production is back-to-back in leading order QCD due to momentum conservation, so the away-side jet partner opposite to a high-$\pt$ particle is likely to have a long path-length of the QCD medium to traverse, maximizing its probability of interaction with the medium. The complementary combination of single and dihadron measurements at high $p_T$ has been very useful at RHIC in constraining the magnitude of geometrical dependence of energy loss~\cite{Jacobs:2004qv}.

Particle angular correlations have been extensively studied at both RHIC and LHC. Two classes of particles are defined: {\it trigger} and {\it associated} particles and the azimuthal angular difference, $\Delta\phi$ and/or the pseudorapidity difference, $\Delta\eta$ are measured for all trigger-associated pairs.  The detector acceptance is constructed from trigger-associated pairs where the trigger and associated particles are taken from different events.  The pair distribution from real events is divided by this acceptance to make a correlation function:
\begin{equation}
C\left(\Delta\phi,\Delta\eta\right) = \frac{d^2N/d\Delta\phi d\Delta\eta}{{\rm Acc}\left(\Delta\phi,\Delta\eta\right)}\,.
\label{eq:corr}
\end{equation}
It is essentially the trigger-associated particle pair density in $\dphi\times\deta$~\cite{Xu:2013sua}.

Typically two classes of correlations are considered.  The first class is the correlations involving a small number of particles, such as the jet-like correlations.  The second class is the entire event-wise correlations.  This class includes correlations between each of the particles and the reaction plane of the collision~\cite{Ollitrault:1992bk}.  The second class of correlations are often described by the Fourier coefficients~\cite{Voloshin:1994mz}.  Showing only the term involving the second harmonic, $v_2$, the event-wise correlations can be described by
\begin{equation}
C_{\rm flow}\left(\Delta\phi,\Delta\eta\right) = B \left( 1 + 2\langle v_{2}^{trig}v_{2}^{assoc}\rangle\cos2\Delta\phi \right)\,.
\label{eq:ewcorr}
\end{equation}
Other harmonics except $v_2$ exist, and will be discussed in Sec.~\ref{sec:vn}. Obtaining $B$ is discussed below.  After subtraction of the terms in Eq.~(\ref{eq:ewcorr}), called flow background, the correlation is thought of in terms of the first class, in particular jet-like correlations. The primary interest is the correlation per jet or trigger particle, so the correlation functions are normalized by the number of trigger particles, and the single particle efficiency is corrected for associated particles. 

The subtraction of the event-wise flow correlation background involves determination of the elliptic flow parameters $v_2$ and the normalization $B$. The $v_2$ parameters are measured by various methods~\cite{Poskanzer:1998yz}, independent of the trigger-associated correlations. The two-particle cumulant method explicitly includes the $v_2$ fluctuation effect in $\langle v_{2}^{trig}v_{2}^{assoc}\rangle$ (see more discussion in Sec.~\ref{sec:vn}). The range of the $v_2$ results from the various methods is often treated as part of the systematic uncertainty in background subtraction. An {\em ad hoc} normalization procedure is used to determine $B$ such that the resultant correlation signal is zero at minimum. This is referred to in literature as Zero Yield At Minimum (ZYAM)~\cite{Adams:2005ph,Ajitanand:2005jj}. It should be noted that ZYAM-background is not a calculation of the combinatoric background level, but an estimate of it which depends to some degree on the shape of the correlated signal to be measured.  

At high $p_T$ the trigger particle can be thought of as a proxy for the jet.  It carries most of the jet momentum~\cite{Boca:1990rh,Kopeliovich:2012sc}. The small-angle trigger-associated correlations show very little variation from \pp, \dau\ to central Au+Au collisions~\cite{Adams:2006yt,Abelev:2009af,Agakishiev:2010ur}, supporting the ``surface emission" hypothesis discussed above. This is consistent with in-vacuum fragmentation of hard-scattered partons emitted from the surface region. On the away-side ($\dphi>\pi/2$) of the trigger particle, the correlation is found to be almost completely suppressed at relatively high $\pt$~\cite{Adler:2002tq}. Such suppressions are not observed in $d$+Au collisions~\cite{Adler:2003ii,Adams:2003im}, indicating jet-quenching is primarily a final-state effect of jet-medium interactions. More detailed discussions of high-$p_T$ jet correlations can be found elsewhere~\cite{Jacobs:2004qv}.

This review concentrates on two novel phenomena that are observed at moderate $p_T$. First, the away-side correlated particles are found to be enhanced in heavy-ion collisions relative to \pp\ and \dau\ at low $\pt$, and in the intermediate $\pt$ range, the correlation function is modified so as to have a double-peaked structure away from $\dphi=\pi$. Second, the near-side correlated particles at small $\dphi$ are found to extend to large pseudorapidities--this is called the ``ridge'' correlation. 
\section{The Away-Side Double-Peak}\label{sec:cone}
\begin{figure}[hbt]
\begin{center}
\includegraphics[width=0.45\textwidth]{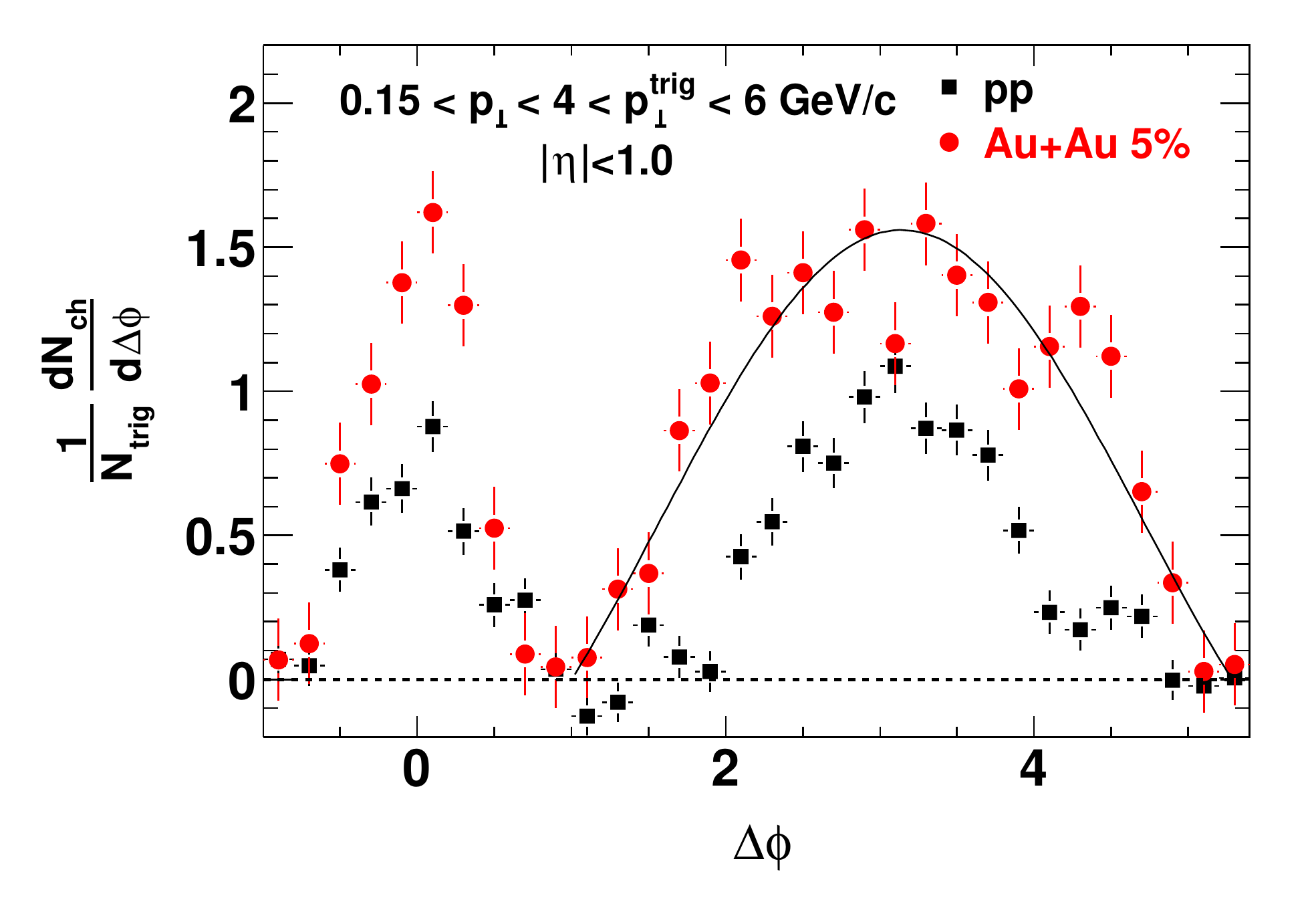}
\includegraphics[width=0.45\textwidth]{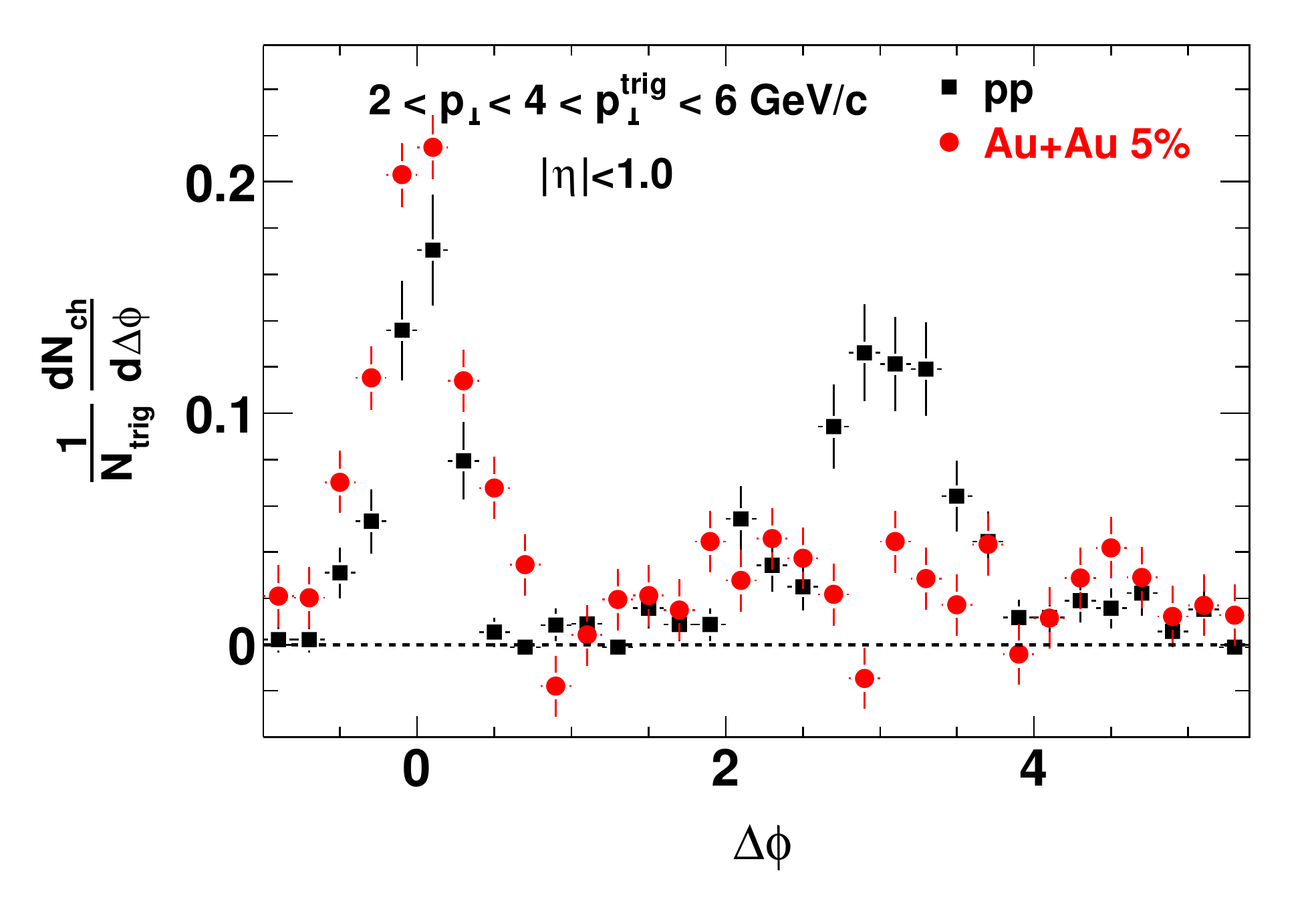}
\end{center}
\caption{Elliptic flow background subtracted correlated yields per trigger in \pp\ (black squares) and  central Au+Au (red circles) collisions.  In both plots the triggers have 4.0$<\ptt<$6~GeV/$c$.  In the left plot associated particles are from 0.15$<\pta<$4.0~GeV/$c$ and in the right from 2.0$<\pta<$4.0~GeV/$c$. From STAR~\cite{Adams:2005ph}.}
\label{fig:STAR_PRL95_Fig1ab}
\end{figure}

\begin{figure}[hbt]
\begin{minipage}{0.6\textwidth}
\begin{center}
\includegraphics[width=\textwidth]{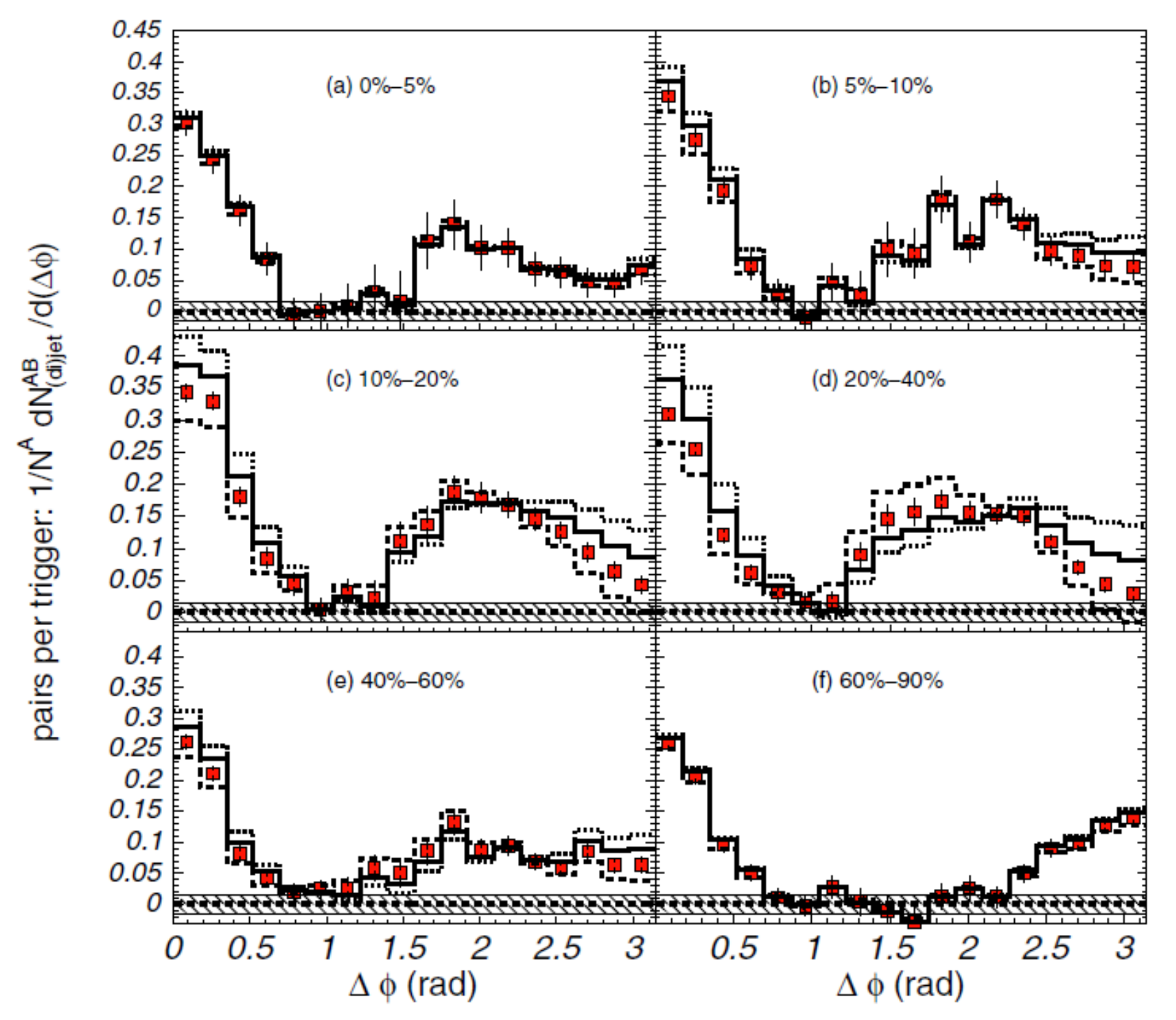}
\end{center}
\end{minipage}
\begin{minipage}{0.4\textwidth}
\caption{Elliptic flow background subtracted correlated yields per trigger in Au+Au collisions of various centralities. The shaded bands indicate the systematic error associated with ZYAM normalization. The dashed (solid) curves are the distributions that would result from increasing (decreasing) $\langle v_2^Av_2^B\rangle$ by one unit of the systematic error; the dotted curve would result from decreasing by two units. From PHENIX~\cite{Adler:2005ee}.}
\label{fig:PHENIX_PRL97}
\end{minipage}
\end{figure}

Figure~\ref{fig:STAR_PRL95_Fig1ab} shows a comparison of jet correlations in \pp\ and central Au+Au collisions by the STAR experiment~\cite{Adams:2005ph}. Jet-like correlations in \pp\ collisions are well measured at $\sqrt{s}$~=~200~GeV~\cite{Adler:2006sc}; The near- and away-side correlations are both approximately Gaussian. The away-side correlations in central heavy-ion collisions, however, are distinctly different from Gaussian. The correlations are significantly broadened, and for intermediate $\pt$ range, even double-peaked~\cite{Adams:2005ph,Adler:2005ee,Adare:2006nr,Adare:2008ae,Aggarwal:2010rf}. Figure~\ref{fig:PHENIX_PRL97} shows the PHENIX results of dihadron correlations for trigger and associated $\pt$ ranges of $2.5<\ptt<4.0$~GeV/$c$ and $1.0<\pta<2.5$~GeV/$c$, respectively~\cite{Adler:2005ee}.  The away-side peak is displaced from $\dphi=\pi$, and displayed over $0<\dphi<2\pi$, the away-side correlation is double-peaked with maxima at $\Delta\phi\approx 2\pi/3$ and $4\pi/3$. PHENIX~\cite{Afanasiev:2007wi} further showed that the double-peak structure has a particle composition similar to the bulk matter. Broad away-side correlations have also been observed in low energy collisions at SPS~\cite{Adamova:2009ah}.  As the trigger particle $\pt$ increases the double-peak becomes 
weaker~\cite{Adare:2008ae}.

The broadened and double-peaked away-side correlations were initially interpreted as possible evidence of Mach-cone shock waves (discussed in Sec.~\ref{sec:machcone}) and/or deflected jets by the QGP medium flow (discussed in Sec.~\ref{sec:deflected}). However, only the elliptic flow $v_2$ background is subtracted in the correlations shown in Figs.~\ref{fig:STAR_PRL95_Fig1ab} and~\ref{fig:PHENIX_PRL97}. As will be discussed in Sec.~\ref{sec:vn}, the away-side broadened and double-peaked correlations may be mainly caused by an additional background of triangular flow, $v_3$. 

\section{The Near-Side Ridge}\label{sec:ridge}
\subsection{The semi-hard ridge}
\begin{figure}[hbt]
\begin{center}
\includegraphics[width=0.45\textwidth]{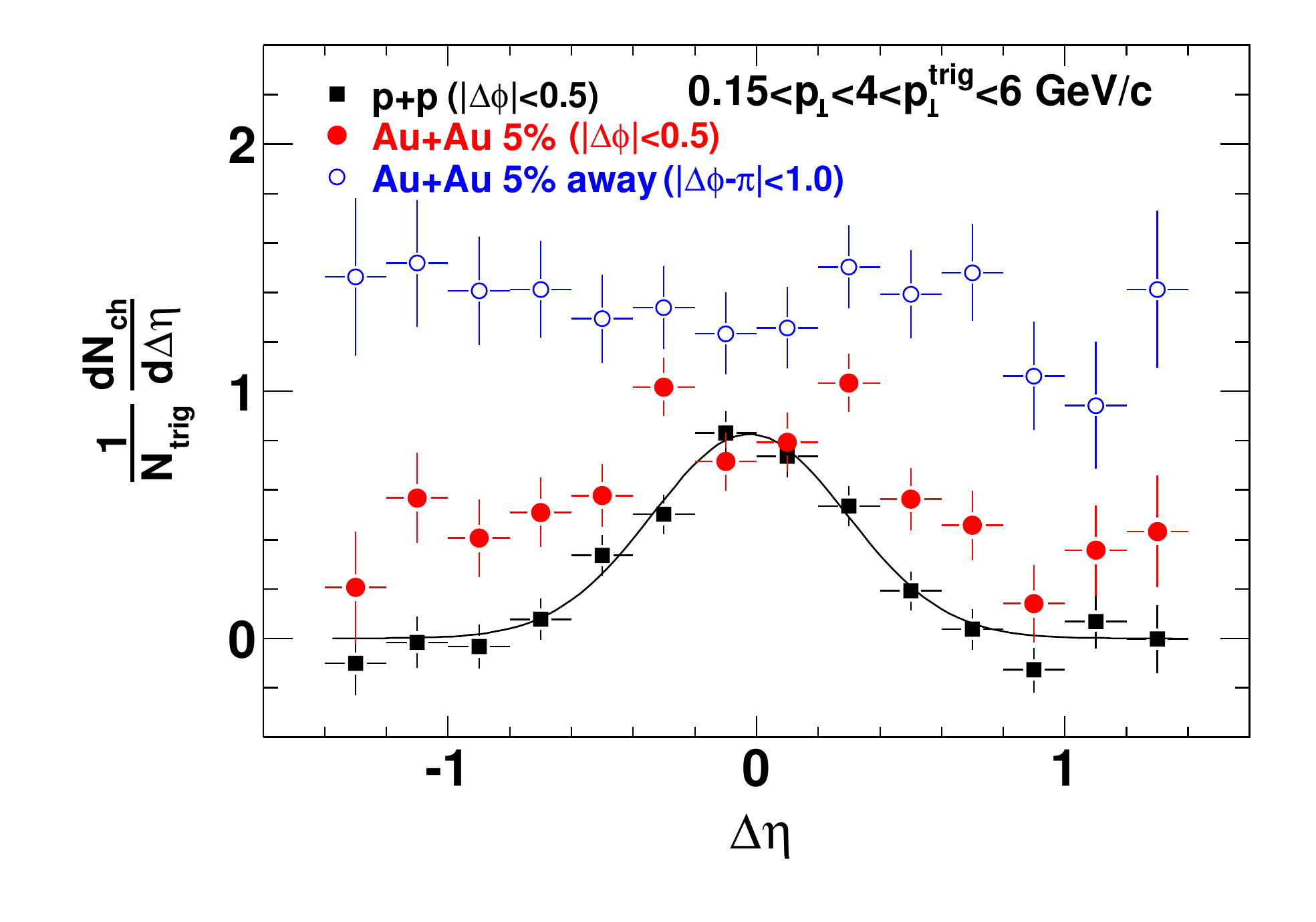}
\includegraphics[width=0.45\textwidth]{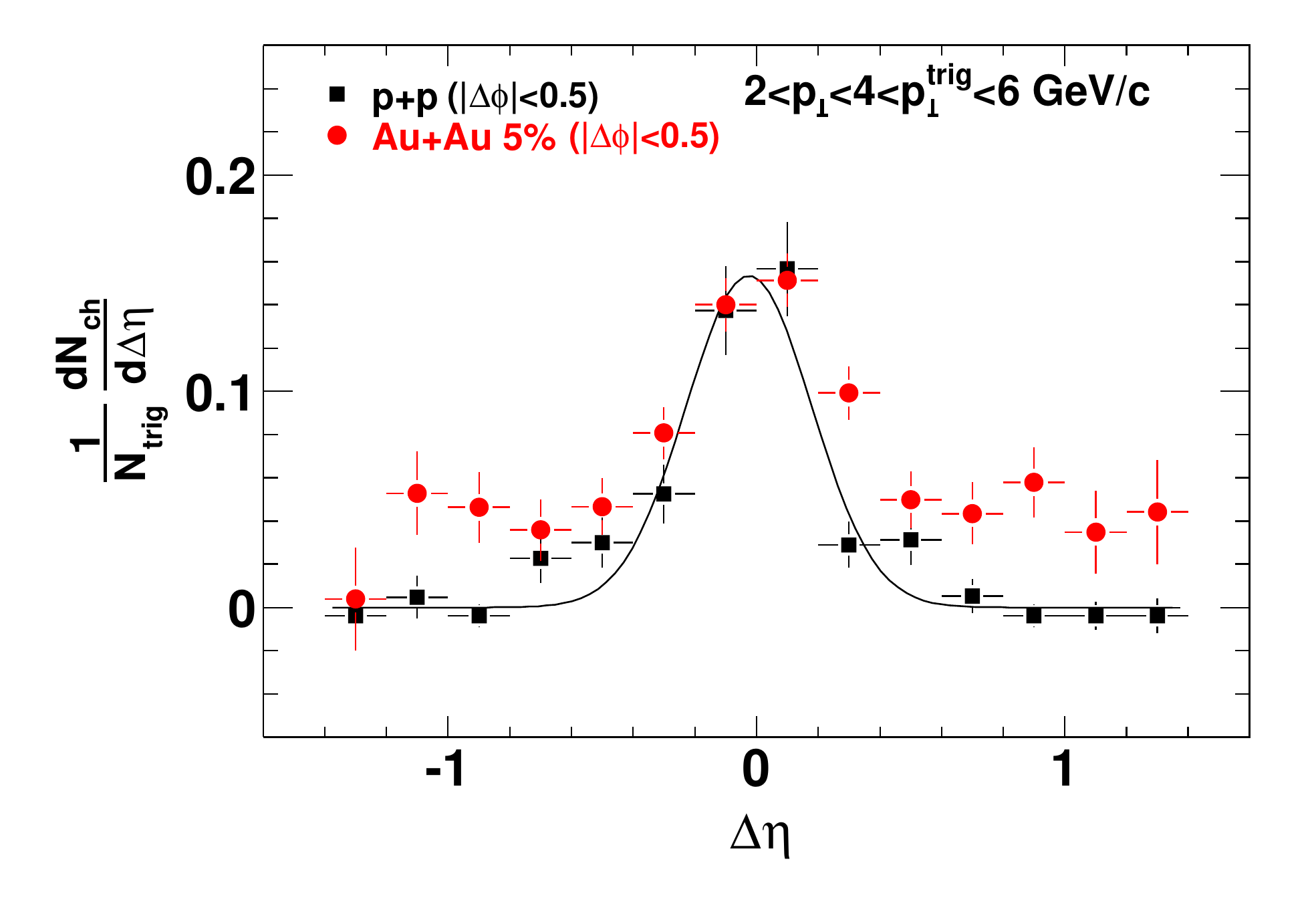}
\end{center}
\caption{ZYAM-background subtracted dihadron $\deta$ correlations in \pp\ and central Au+Au collisions at $\snn=200$~GeV. The near-side correlations are shown in black squares and red circles  for \pp\ and Au+Au, respectively. In both panels the triggers have 4.0$<\ptt<$6~GeV/$c$.  In the left panel associated particles are from 0.15$<\pta<$4.0~GeV/$c$ and in the right from 2.0$<\pta<$4.0~GeV/$c$. The left panel also shows the away-side correlations for Au+Au. From STAR~\cite{Adams:2005ph}.}
\label{fig:STAR_PRL95_Fig1}
\end{figure}

In the same kinematic region as the away side double-peak structure, a small $\Delta\phi$ large $\Delta\eta$ correlation is observed, the ridge. The ridge is first observed in Au+Au collisions at $\snn=200$~\gev\ at RHIC~\cite{Adams:2005ph}. Dihadron correlations are analyzed in hadron pair azimuthal and pseudorapidity differences $(\dphi,\deta)$. 
The combinatoric background is subtracted via ZYAM. Figure~\ref{fig:STAR_PRL95_Fig1} shows the dihadron correlation results from STAR with a trigger particle of $4<\ptt<6$~\gev~\cite{Adams:2005ph}. It is found that the ZYAM-background subtracted $dN/d\deta$ correlation signal is broader in Au+Au collisions than in elementary \pp\ collisions. At relatively large associated $\pt$, the correlation signal appears to be atop an approximately uniform pedestal in $\deta$. 

\begin{figure}[hbt]
\begin{center}
\includegraphics[width=0.45\textwidth]{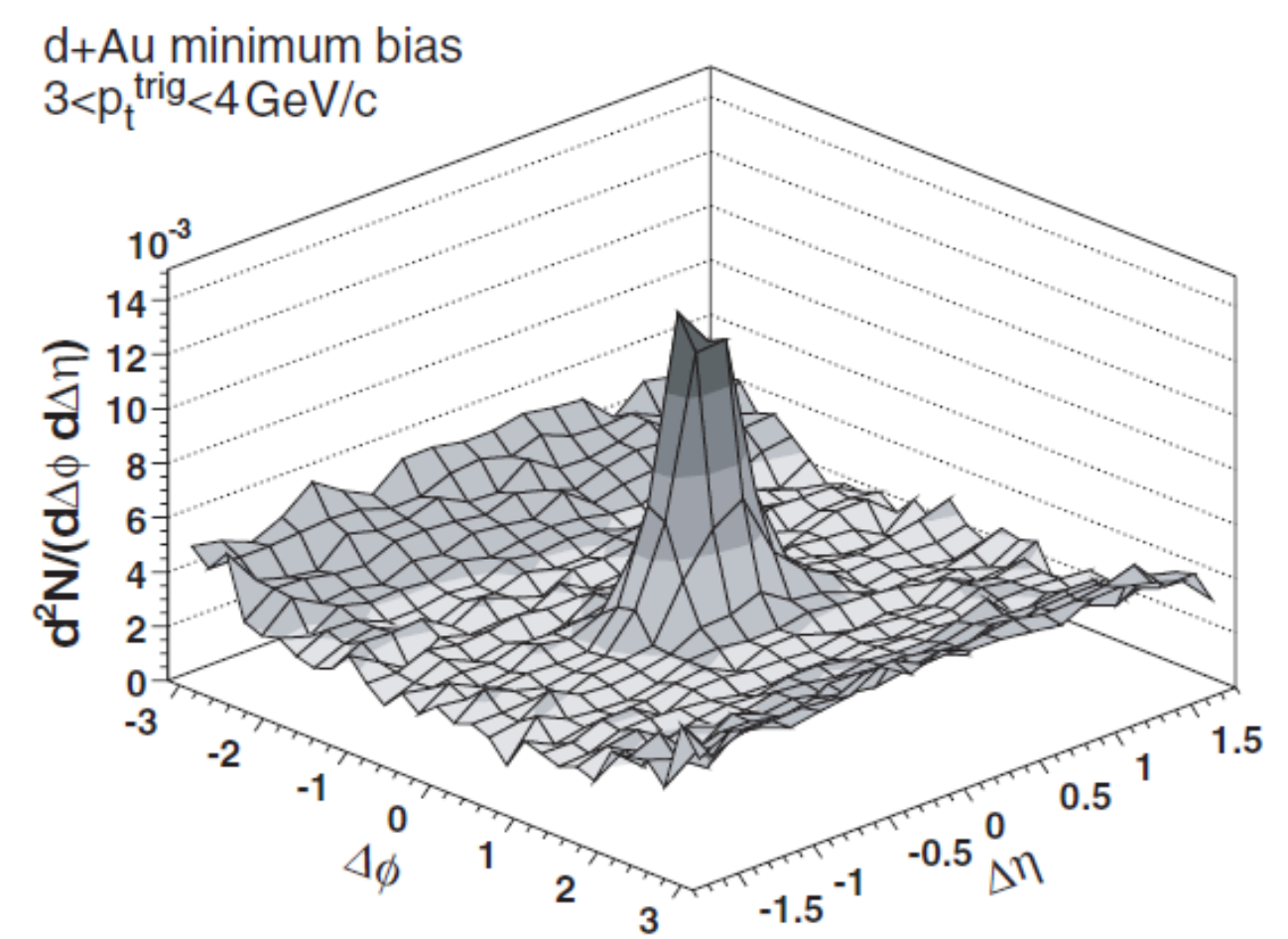}
\includegraphics[width=0.45\textwidth]{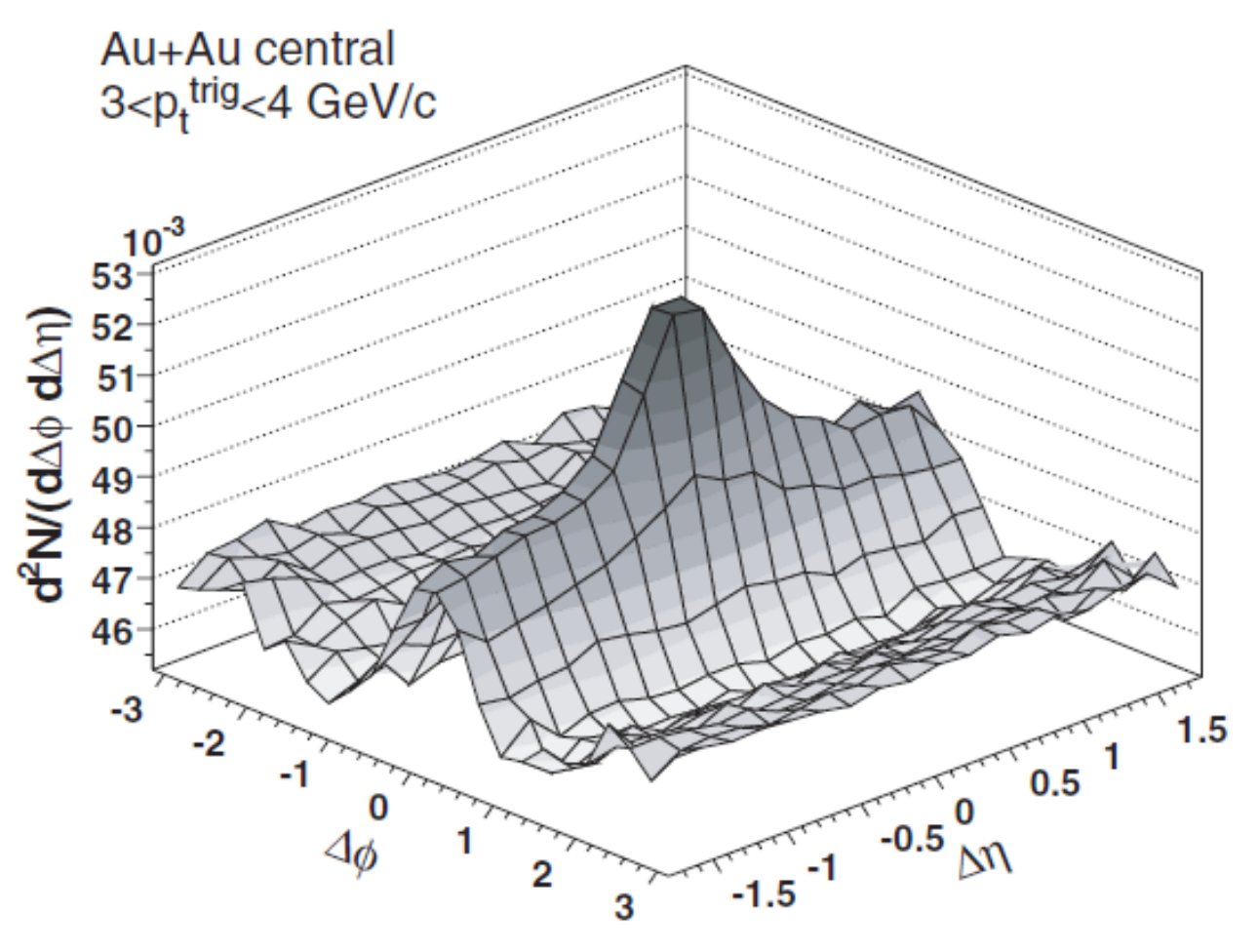}
\end{center}
\caption{Two-dimensional dihadron correlation functions in $(\deta,\dphi)$ in minimum bias \dau\ (left panel) and 12\% central Au+Au collisions (right panel). The trigger and associated particle $\pt$ ranges are $3<\ptt<4$~\gev\ and 2~\gev$<\pta<\ptt$, respectively. From STAR~\cite{Abelev:2009af}.}
\label{fig:STAR_PRC80_Fig1}
\end{figure}

It should be noted that ZYAM-background is only an {\it ad hoc} solution in lack of better understanding of the correlation background. It is conceivable that the real background is lower when near- and away-side jet-correlation peaks start to overlap, which is likely at low $\pt$ where the peaks are broad. The qualitative observation of the existence of the ridge is, however, robust.
The ZYAM-normalization approach is visualized more clearly in Fig.~\ref{fig:STAR_PRC80_Fig1}~\cite{Abelev:2009af}, where the approximately uniform $\deta$ pedestal is evident on the near-side of the trigger particle ($\dphi\approx 0$) above the minimum valley in $\dphi$. 

\begin{figure}[hbt]
\begin{minipage}{0.5\textwidth}
\begin{center}
\includegraphics[width=\textwidth]{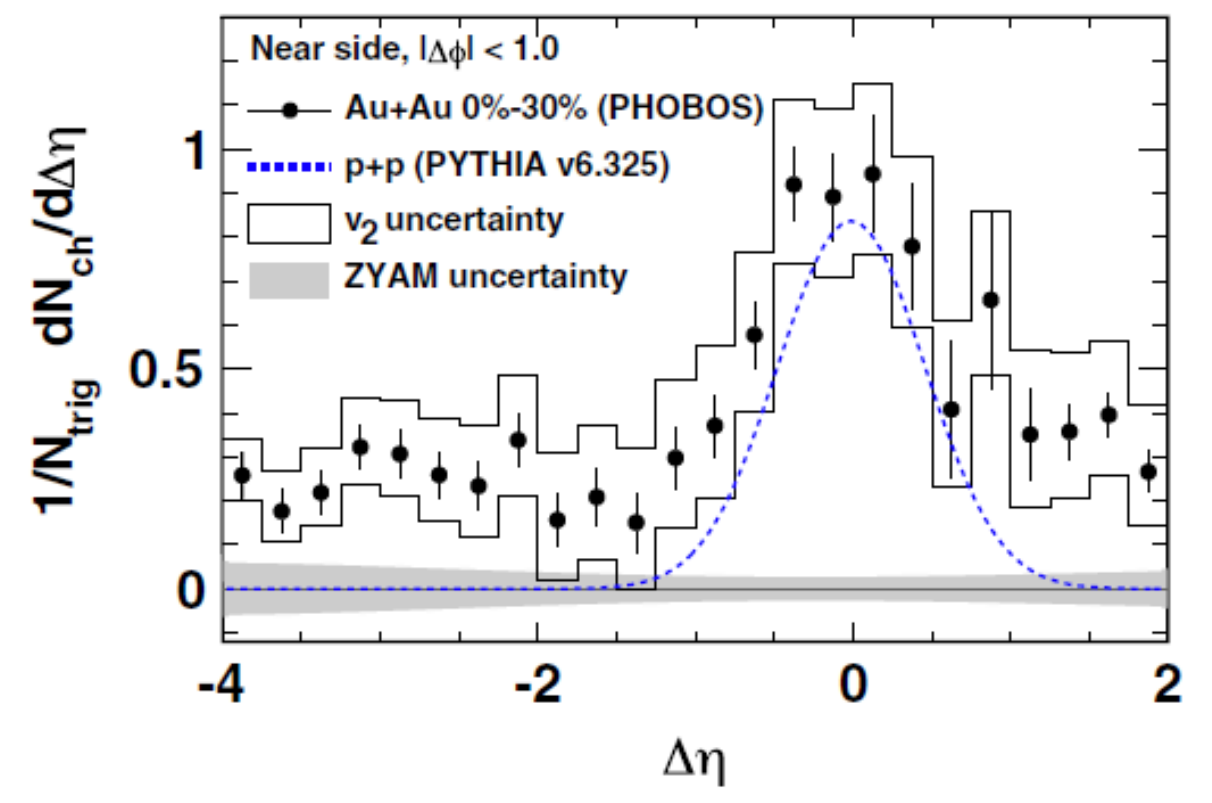}
\end{center}
\end{minipage}
\begin{minipage}{0.5\textwidth}
\caption{Near-side yield integrated over $|\Delta\phi|<1$ for 0-30\% Au+Au compared to PYTHIA \pp\ (dashed line) as a function of $\Delta\eta$. The trigger particle is at $\pt>2.5$~\gev\ and the associated particles are from essentially all $\pt$. Bands around the data points represent the uncertainty from flow subtraction. The error on the ZYAM procedure is shown as a gray band at zero. All systematic uncertainties are 90\% confidence level. From PHOBOS~\cite{Alver:2009id}.}
\label{fig:PHOBOS_PRL104}
\end{minipage}
\end{figure}

The ridge is observed to extend to several units in $\deta$ by PHOBOS~\cite{Alver:2009id} and STAR~\cite{Wang:2008zzh}. Figure~\ref{fig:PHOBOS_PRL104} shows the near-side $\deta$ correlation function with a $\ptt>2.5$~\gev\ trigger particle from PHOBOS~\cite{Alver:2009id}. An approximately uniform ridge is observed to extend to at least 4 units in $\deta$.

The ridge magnitude increases with the event centrality. In peripheral collisions no ridge is observable~\cite{Adams:2005ph,Abelev:2009af}. The ridge is found to be approximately independent of the trigger particle $\pt$~\cite{Abelev:2009af}. 
Similarly to the double-peak away-side structure, the ridge properties appear to be more similar to the bulk than jets. The $\pt$ spectrum of the correlated ridge particles is softer than that of the jet particles (i.e.~those above the uniform ridge pedestal) and  only slightly harder than the inclusive hadrons~\cite{Abelev:2009af}. The (anti-)baryon content in the ridge is increased relative to \pp\ collisions, similar to inclusive hadrons, not to jet-correlated hadrons~\cite{Nattrass:2010jn}. 

In order to investigate the origin of the ridge, a three-particle correlation analysis~\cite{Abelev:2009jv} is carried out  where two associated particles of $1<\pta<2$~\gev\ are correlated to a trigger particle of $\ptt>3$~\gev. It is found that the correlations of ridge particles are uniform not only with respect to the trigger particle but also between themselves event-by-event. In addition, the production of the ridge appears to be uncorrelated to the presence of the narrow jet-like component. 

The ridge in heavy-ion collisions is observed after subtraction of the $v_2$ background. As will be discussed in Sec.~\ref{sec:vn}, the ridge may be mainly due to higher order hydrodynamic flow harmonics, particularly $v_3$ that have been overlooked until recently. The double-peak away-side correlation and the near-side ridge correlation may, therefore, to a large extent be manifestations of the same underlying physics of high-order hydrodynamic flows. 

\subsection{The soft ridge}
\begin{figure}[hbt]
\begin{minipage}{0.45\textwidth}
\begin{center}
\includegraphics[width=\textwidth]{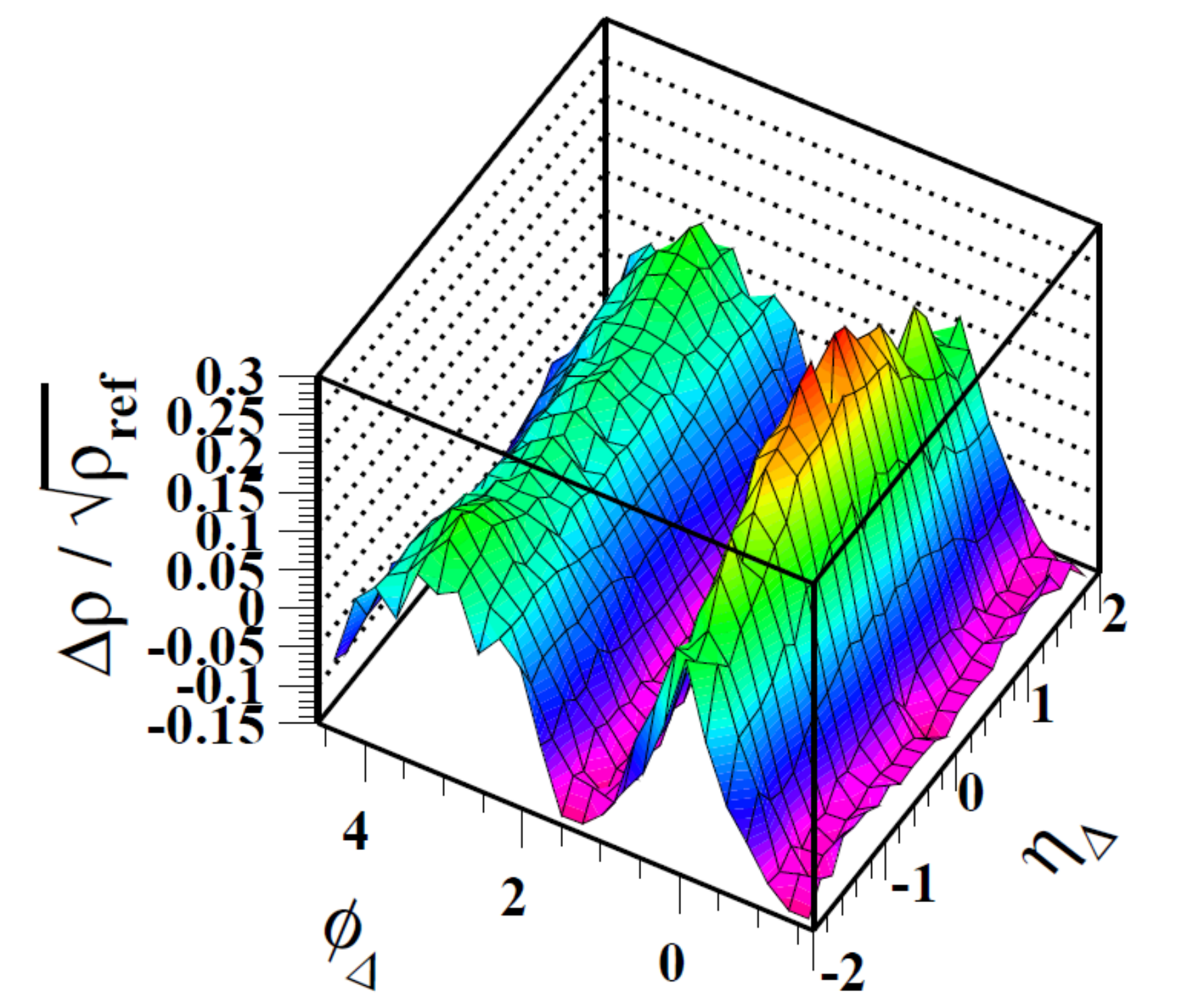}
\end{center}
\end{minipage}
\begin{minipage}{0.55\textwidth}
\caption{Untriggered (no $\pt$ cut) two-particle correlations as a function of $\Delta\eta$ and $\Delta\phi$ in top 5\% central Au+Au collisions at 200 GeV. From STAR~\cite{Agakishiev:2011pe}.}
\label{fig:STAR_PRC86_Fig1d}
\end{minipage}
\end{figure}

Additionally, a long-range pseudorapidity correlation is also observed in low-$\pt$ dihadron correlations without a high-$\pt$ trigger particle~\cite{Adams:2004gp,Adams:2004pa}. This is often referred to as ``untriggered'' correlation in contrast to the high-$\pt$ ``triggered'' correlation. In these correlations, the same ratio of the number of real to mixed pairs is constructed. 
The untriggered correlation in central Au+Au collisions at 200 GeV is shown in Fig.~\ref{fig:STAR_PRC86_Fig1d}. It shows a broadened $\deta$ distribution, which is often referred to as the untriggered or soft ridge. Without the high-$p_T$ trigger particle jet-like features are greatly reduced and also broadened.

STAR~\cite{Agakishiev:2011pe} analyzed the untriggered correlation with a fit model composing of a near-side two-dimensional Gaussian, a negative dipole, and a quadrupole: 
\begin{equation}
C=A_0+A_{ns}\exp\left[-\frac{1}{2}\left(\frac{\dphi}{\sigma_{\dphi}}\right)^2-\frac{1}{2}\left(\frac{\deta}{\sigma_{\deta}}\right)^2\right]-A_D\cos\dphi+A_Q\cos2\dphi\,.\label{eq:estruct}
\end{equation}
The fitted Gaussian and dipole magnitudes show a strong increase with the collision centrality exhibiting a shape transition approximately at the 40\% centrality percentile~\cite{Agakishiev:2011pe} (smaller percentile means more central collisions). The $\deta$ width of the two-dimensional Gaussian increases significantly with centrality while the $\dphi$ width decreases modestly~\cite{Agakishiev:2011pe}. 

The fitted quadrupole does not always coincide with the elliptic flow parameter measured by other methods. For central collisions, it is outside the range between the two- and four-particle cumulant measurements~\cite{Kettler:2009zz}. This suggests that the mathematical description of the correlation function by Eq.~(\ref{eq:estruct}) may not faithfully separate the two contributions of jet-like and event-wise correlations. 
One component that is missing from Eq.~(\ref{eq:estruct}) is a sextupole (triangular anisotropy). Although the untriggered correlation data cannot discriminate between Eq.~(\ref{eq:estruct}) and a model including an additional sextupole~\cite{Agakishiev:2011pe}, other measurements and theoretical studies suggest that higher-order harmonics should be present in untriggered particle correlations. 
\section{Higher Order Flow Harmonics}\label{sec:vn}
\begin{figure}
\begin{minipage}{0.5\textwidth}
\begin{center}
\includegraphics[width=\textwidth]{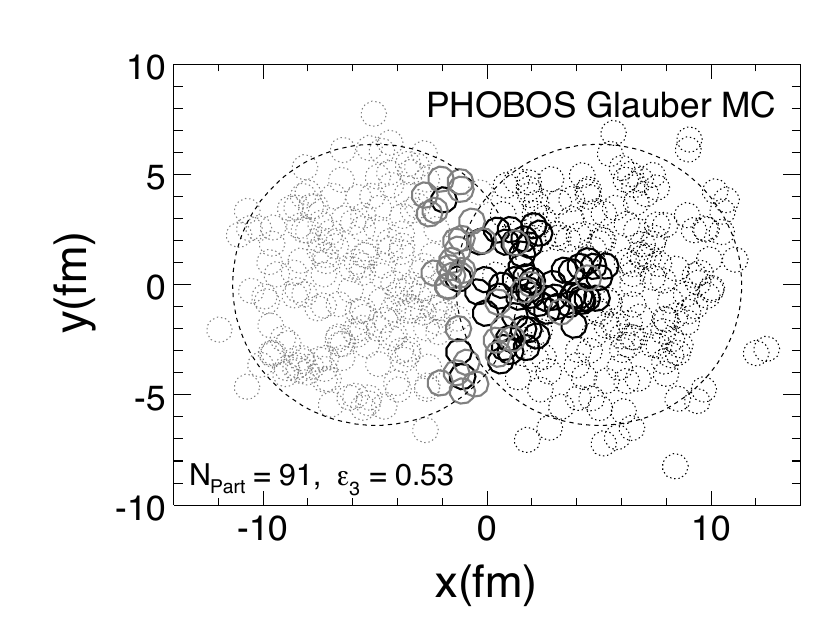}
\end{center}
\end{minipage}
\begin{minipage}{0.5\textwidth}
\caption{Transverse positions of nucleons from a Au+Au collision generated in a Glauber Monte Carlo simulation. The outlines of the two nuclei are shown as circles.  The non-interacting nucleons are shown as light small circles and the participating nucleons are shown in black. From Ref.~\cite{Alver:2010gr}.}
\label{fig:glauber_projection}
\end{minipage}
\end{figure}

In most of the correlations discussed in the previous sections, only background from $v_2$ 
is subtracted from the correlation functions, as in Eq.~(\ref{eq:ewcorr}). This was expected to be the dominant background contribution due to the elliptic shape of the overlap region between the two colliding nuclei. Higher order even harmonics were thought to be small compared to $v_2$ and odd harmonics were thought to be zero because of the symmetry of the system.
It has not been realized until recently~\cite{Mishra:2007tw,Alver:2010gr} that such symmetry does not exist on event-by-event basis because of quantum fluctuations in initial nucleon distributions in nuclei and that these fluctuations have observable consequences in the final state.  As an example, Fig.~\ref{fig:glauber_projection} depicts the initial positions of nucleons in an event generated with a Glauber Monte Carlo where it is clear there is nonzero triangularity~\cite{Alver:2010gr}. 
Event-by-event, the single particle azimuthal distribution can be described by
\begin{equation}
dN/d\phi\propto 1+\sum_{n=1}^{\infty}2v_n\cos n(\phi-\psi_n)\,,\label{eq:vn}
\end{equation}
where $\psi_n$ is the $n^{\rm th}$ order harmonic plane. The $v_n$ anisotropies result in a modulation of two-particle distributions in relative azimuthal angle, 
\begin{equation}
dN/d\dphi\propto 1+\sum_{n=1}^{\infty}2\langle v_n\trig v_n\asso\rangle\cos n\dphi\,,
\label{eq:vn2}
\end{equation}
including all orders of $n$ in contrast to Eq.~(\ref{eq:ewcorr}).
Thus, there can be finite contributions to correlations from odd harmonic anisotropies, such as $\langle v_3^{(t)}v_3^{(a)}\rangle\cos3\dphi$. Monte Carlo models including the effects of event-by-event fluctuations are able to reproduce the qualitative features of the double-peak and ridge correlations~\cite{Ma:2006fm,Xu:2010du,Xu:2011fe,Xu:2011jm,Takahashi:2009na}, explaining both phenomena in terms of initial state fluctuations.
In fact, a third harmonic component has always been present in all experimental dihadron correlation measurements, including untriggered low-$\pt$ correlations~\cite{Abelev:2008un}. It is not attributed to any flow contributions until the  realization that odd-harmonics can exist due to event-by-event fluctuations of the initial distributions of energy density.

\subsection{Measurements of odd harmonics}
Qualitatively, the measured two-particle correlations are consistent with a background contribution from odd-harmonics, particularly $v_3$. The $v_3$ term in Eq.~(\ref{eq:vn2}) give a three-peak structure, the peak at $\dphi=0$ contributes to the near-side ridge, and the peaks at $\dphi=2\pi/3$ and $4\pi/3$ are the away-side double-peak structure. Quantitatively, it is a much harder question to answer whether or not the measured correlations are entirely due to hydrodynamic flow. This is because  flow measurements are contaminated by non-hydrodynamical sources.

There are two main methods to measure anisotropic flow. One is the event-plane method where the particle of interest is correlated with a reconstructed event plane. The other is the two-particle cumulant method, where the Fourier coefficients of two-particle correlations are evaluated. These two methods are connected because the event plane is reconstructed using final-state particle momenta, not initial state geometry configuration. Correlation of the particle of interest to the event plane is therefore made of many two-particle correlations. The measured anisotropies are therefore a net effect of single particle hydrodynamic flow and intrinsic two- and multi-particle correlations. Those few-body correlations are unrelated to the flow event plane and are often called nonflow correlations~\cite{Borghini:2000cm,Wang:2008gp}. Nonflow correlations are typically of short range. Part of the nonflow contributions comes from jet correlations. Other nonflow contributions may include resonance decays, charge conservation effects~\cite{Bozek:2012en}, the color glass condensate (discussed in Sec.~\ref{sec:cgc}), and other physics processes. 

The four-particle cumulant method has also been used for flow measurements. Four-particle cumulant is insensitive to nonflow contributions because of suppression by large multiplicity to the third power~\cite{Borghini:2000sa,Borghini:2001vi}. However, flow fluctuation effects are negative in four-particle cumulant~\cite{Borghini:2001vi}, making it less suitable for jet-correlation background subtraction, to which the contribution of flow fluctuations is positive. 

To reduce short-range nonflow correlations, in particular the near-side intra-jet correlations, an pseudorapidity $\eta$ gap is often applied between the two particles in two-particle anisotropy measurements. Because the away-side jet partner has a broad distribution in $\eta$ relative to the near-side trigger particle, the away-side jet correlations are not removed by this method. Recently it was suggested that there might be an $\deta$-dependent component in flow fluctuations~\cite{Bozek:2010vz,Petersen:2011fp,Xiao:2012uw}. An $\eta$ gap might therefore not only reduce nonflow correlations but also flow fluctuations. 

To measure flow correlations with two-particle cumulant, one would like to form pairs with no or minimal contributions from nonflow correlations. In particular, to measure background to jet-like correlations, one would like to have pairs without contributions from jets~\cite{Wang:2009af}. One viable way is to use low $\pt$ particles which minimize, but does not eliminate, the jet contributions. In this approach, one correlates the particle of interest to a reference particle at low enough $\pt$, and obtain the Fourier coefficients of the correlations, $\Vn{\pt,\ptref}$. One can also obtain the Fourier coefficients of correlations between two reference particles, $\Vn{\ptref,\ptref}$. Assuming negligible nonflow contributions to these Fourier coefficients, one obtain the single particle flow anisotropies of the particle of interest by
\begin{equation}
v_n(\pt)=\Vn{\pt,\ptref}\left/\sqrt{\Vn{\ptref,\ptref}}\right.\,.\label{eq:vn_fact}
\end{equation}
Note that the quantities in Eq.~(\ref{eq:vn_fact}) all include flow fluctuation effects. A clearer (but a bit cumbersome) expression of Eq.~(\ref{eq:vn_fact}), with the assumption of vanishing nonflow, should be
\begin{equation}
\sqrt{\langle v_n^2(\pt)\rangle}=\langle v_n(\pt)v_n(\ptref)\rangle\left/\sqrt{\langle v_n^2(\ptref)\rangle}\right.\,.\label{eq:vn_fact_fluc}
\end{equation}
In general, fluctuations break the equality of Eqs.~(\ref{eq:vn_fact}) and~(\ref{eq:vn_fact_fluc}). Only when fluctuations are linear or small do Eqs.~(\ref{eq:vn_fact}) and~(\ref{eq:vn_fact_fluc}) hold.

\begin{figure}[hbt]
\begin{minipage}{0.45\textwidth}
\begin{center}
\includegraphics[width=\textwidth]{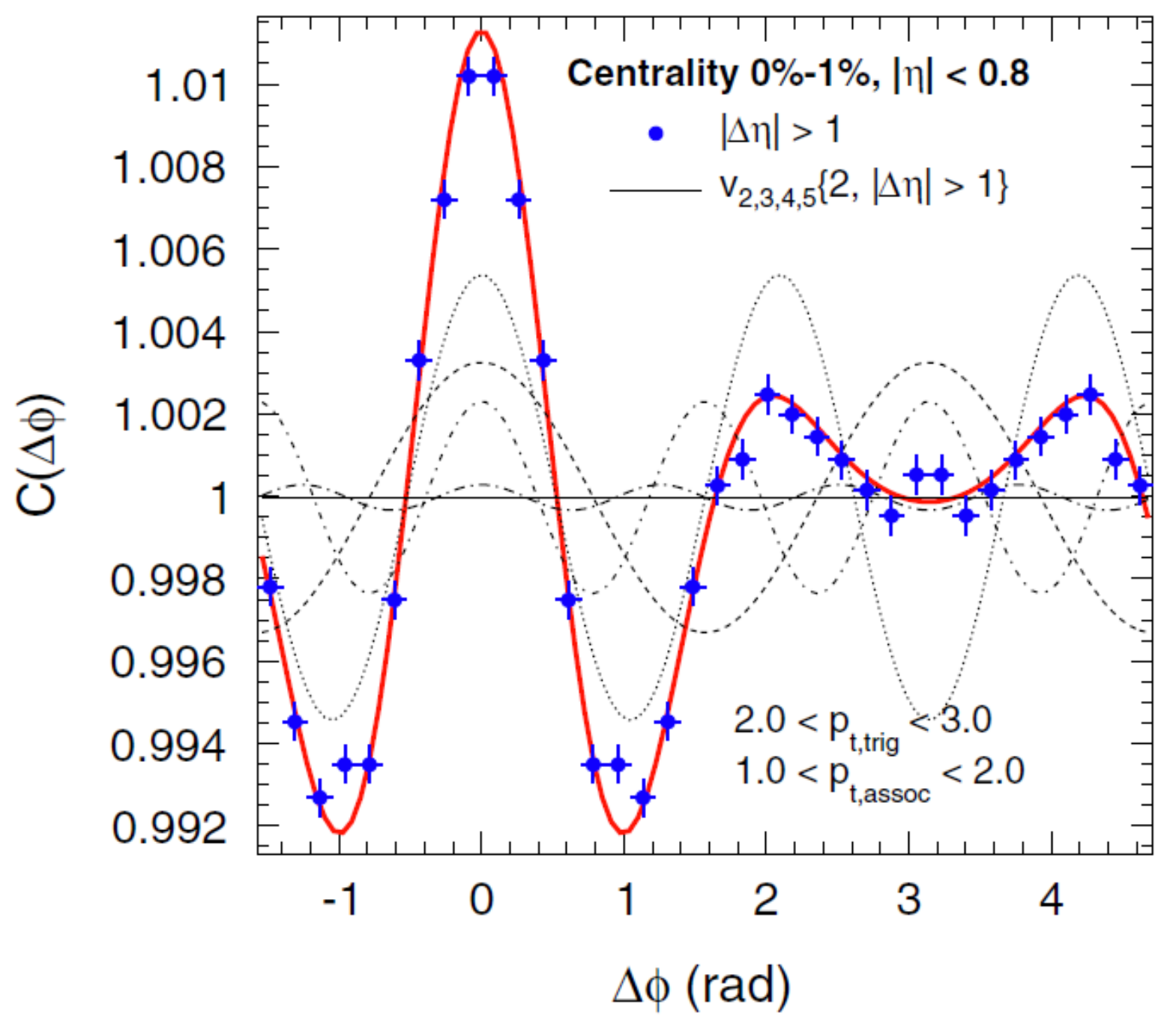}
\end{center}
\end{minipage}
\begin{minipage}{0.55\textwidth}
\caption{The two-particle azimuthal correlation, measured in $0<\dphi<\pi$ and shown symmetrized over $2\pi$, between a trigger particle with $2<\ptt<3$~\gev\ and an associated particle with $1<\pta<2$~\gev\ for top 1\% Pb+Pb collisions at the LHC. The solid red line shows the sum of the measured anisotropic flow Fourier coefficients $v_2$, $v_3$, $v_4$, and $v_5$ (dashed lines). From ALICE~\cite{ALICE:2011ab}.}
\label{fig:ALICE_PRL107_Fig4}
\end{minipage}
\end{figure}

It is shown by ALICE~\cite{ALICE:2011ab} that the two-particle correlations between a trigger particle of $2<\ptt<3$~\gev\ and an associated particle of $1<\pta<2$~\gev\ can be completely described by ``single'' (trigger and associated) particle anisotropies (see Fig.~\ref{fig:ALICE_PRL107_Fig4}). The single particle anisotropies are obtained by two-particle correlations between the particle of interest and a reference particle from the inclusive $\pt$ range of $0<\ptref<5$~\gev. Because the reference particle is not particularly from low $\pt$ but encompassing the trigger and associated $\pt$ ranges, the complete description of the trigger-associated correlation by the single particle anisotropies may not be surprising, even if there may be nonflow contributions in those Fourier coefficients~\cite{Kikola:2011tu}.

If two-particle correlations are only due to single-particle flow, Fourier coefficients of two-particle correlations would equal to the product of the root-mean-square of the single particle flow anisotropies. This is referred to as ``factorization''~\cite{Alver:2010dn,Luzum:2010sp}. In order to quantify the degree of factorization, 
one can expand the two-particle correlations into Fourier series of harmonics,
\begin{equation}
\frac{dN}{d\dphi}\propto 1+\sum_{n=1}^{\infty}2V_n\cos(n\dphi)\,,
\end{equation}
where the coefficients are given by 
\begin{equation}
V_n=\langle\cos(n\dphi)\rangle\,.
\end{equation}
The LHC experiments~\cite{Chatrchyan:2011eka,Aamodt:2011by,Chatrchyan:2012wg,ATLAS:2012at} have all decomposed two-particle $\dphi$ correlation functions (with large $\deta$) into Fourier series. 
The experimental data indicate that this factorization indeed holds to a good approximation at low-intermediate $\pt$ ($\pt<4$~\gev) for all harmonics except the $v_1$. Thus, a global fit using a single $v_n(\pt)$ can describe well the two-particle correlation Fourier coefficients. With increasing $\pt$, the factorization breaks down because of more significant contributions from back-to-back jet-correlations. 

However, it is shown~\cite{Kikola:2011tu} that jet-correlations from PYTHIA 
simulations of \pp\ collisions also approximately factorize. It is further shown~\cite{Kikola:2011tu}, due to the bootstrap nature of the Fourier coefficients, the factorization can hold to a good precision even when there is a significant nonflow contribution to two-particle correlations. 
It is, therefore, clear that good degree of factorization of Fourier coefficients does not necessarily mean they must be entirely hydrodynamic in nature.

All experiments~\cite{ALICE:2011ab,Adare:2011tg,Chatrchyan:2012wg,ATLAS:2012at,Adamczyk:2013waa} have measured $v_3$ and higher harmonics by the event-plane and the two-particle correlation methods. While the theoretical goal was for $v_2$ to be a measure of the azimuthal distribution of single particles with respect to the reaction plane~\cite{Ollitrault:1993ba}, it was later realized that due to event-by-event fluctuations of nucleon distributions and nucleon-nucleon interactions, $v_2$ is measured with respect to the second order  participant plane~\cite{Alver:2008zza}, $\psi_2$. Measurements of $v_n$ at all orders can be made in the same manner with respect to the $n^{\rm th}$ order event plane, $\psi_n$.  The various planes are not necessarily correlated with each other or to the reaction plane. For example, the $v_3$ with respect to the second harmonic event plane is very small~\cite{ALICE:2011ab,Adare:2011tg}, suggesting the magnitude of  $v_3$ is related to geometrical fluctuations rather than geometrical correlations.

\begin{figure}
\begin{center}
\includegraphics[width=0.43\textwidth]{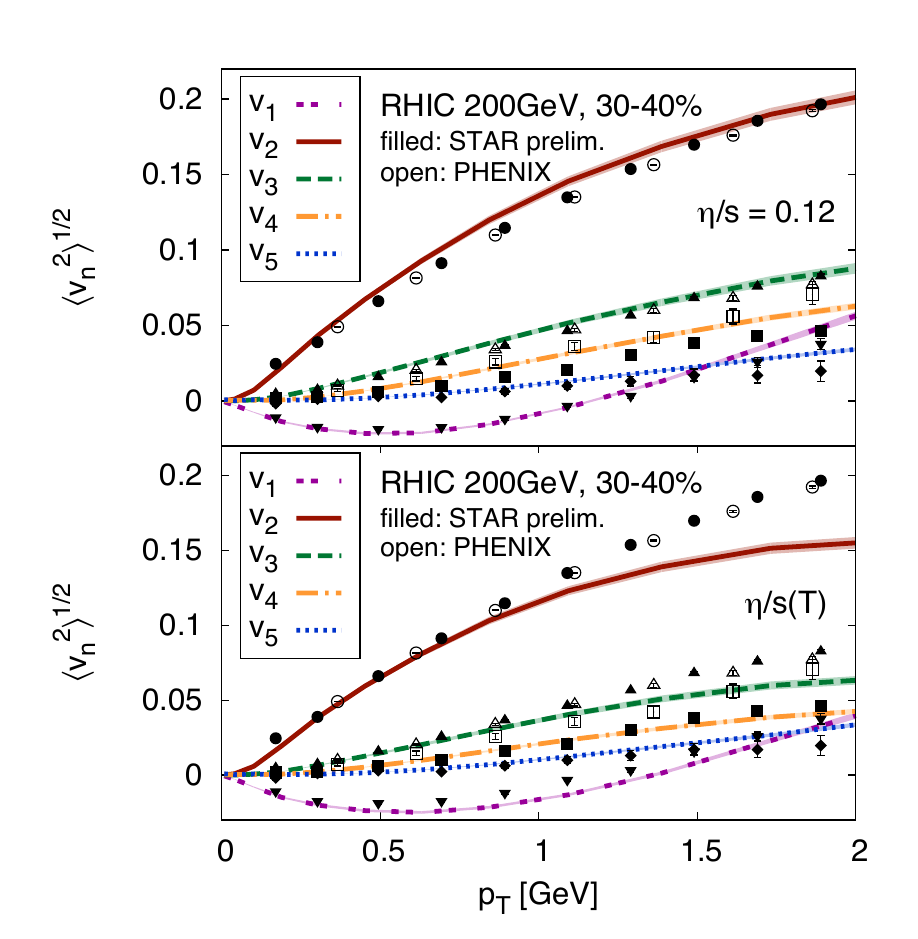}
\includegraphics[width=0.45\textwidth]{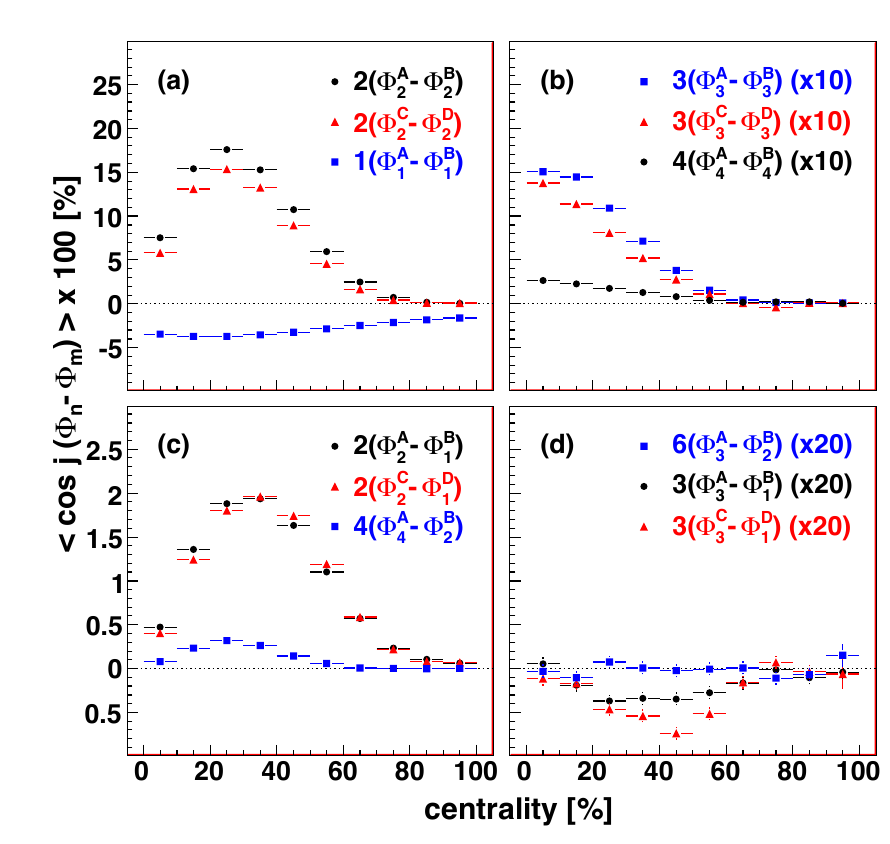}
\end{center}
\caption{(Left panel) $v_n$ results at $\sqrt{s_{NN}}$=200~GeV~\cite{Adare:2011tg,Pandit:2012mq} as compared to theoretical calculations from the IP-Glasma+MUSIC model~\cite{Gale:2012rq}. (Right panel) Raw correlation strength between event planes as measured in various detector combinations as a function of collision centrality from PHENIX~\cite{Adare:2011tg}.  The letter superscripts denote various PHENIX detector subsystems with the corresponding $\eta$ coverage, `A' for RXN North ($1.0<\eta<2.8$), `B' for BBC South ($-3.9<\eta<-3.1$), `C' for MPC North ($3.1<\eta<3.7$), and `D' for MPC South ($-3.7<\eta<-3.1$). The number subscripts denote the order of the event plane.}
\label{fig:phenix_v3}
\end{figure}

Measurements by PHENIX and STAR show the connection between odd harmonics and initial state fluctuations by comparing $v_3$ measured with respect to the third order event plane to hydrodynamic calculations~\cite{Adare:2011tg,Pandit:2012mq}.  This is shown in the left panel of Fig.~\ref{fig:phenix_v3}.  Additional evidence for the connection between $v_n$ and the initial state fluctuations is provided by measurements of the correlations between event planes in both sides of the collision region~\cite{Adare:2011tg} as shown in the right panel of Fig.~\ref{fig:phenix_v3}. The PHENIX measurements of the correlations between $\psi_3$ are from detectors on opposite sides of the collision region  with a pseudorapidity difference of approximately 7 units~\cite{Adare:2011tg}.
 
The $v_3$ measured by STAR~\cite{Adamczyk:2013waa} decreases significantly with the $\eta$-gap between the two particles (or between the particle of interest and the event plane) used in the measurement. This is initially observed in the broad $\deta$ Gaussian of the near-side peak in untriggered dihadron correlations~\cite{Agakishiev:2011pe}. 
There might be tension between the STAR and PHENIX $v_3$ measurements. STAR has measured a significantly smaller midrapidity $v_3$ using event plane from the forward TPC (FTPC), with a $\deta\sim3$, than from the main TPC, $\deta\sim0.6$~\cite{Adamczyk:2013waa}. But the STAR TPC midrapidity result with midrapidity TPC event plane is consistent with the PHENIX midrapidity $v_3$ result measured at forward rapidity ($\deta\sim2$)~\cite{Adamczyk:2013waa,Adare:2011tg}. 
These apparent discrepancies may arise from the various $\eta$ gaps used in the measurements together with the possible event-plane decorrelations over $\deta$~\cite{Xiao:2012uw}.

\subsection{Odd harmonics in hydrodynamic calculations}
Hydrodynamic calculations using initial conditions averaged over many events are used successfully to describe elliptic flow data at RHIC and LHC~\cite{Shen:2010uy}. A single hydrodynamic run is performed for the event class of interest and these are sometimes called ``single-shot" calculations. Event-by-event hydrodynamic calculations, where initial configuration of each individual event is used as the starting point for the hydrodynamic simulation, are needed in order to calculate odd harmonic flow~\cite{Holopainen:2010gz,Qin:2010pf,Schenke:2010rr,Qiu:2011iv}. Hydrodynamic evolutions translate the initial configuration space anisotropy (and fluctuations) into final-state momentum anisotropy~\cite{Qin:2010pf}. 
The importance of event-by-event fluctuations in elliptic flow calculations was in fact realized early-on~\cite{Bhalerao:2006tp,Andrade:2006yh,Alver:2008zza}. The realization of event-by-event triangular geometry is only recent~\cite{Mishra:2007tw,Sorensen:2010zq,Alver:2010gr,Alver:2010dn}. Just like $v_2$ corresponding to the initial elliptical geometry, the final state $v_3$ results from the hydrodynamical response to the triangular initial geometry~\cite{Alver:2010dn,Teaney:2010vd}.

An illustrative demonstration of the hydrodynamic evolution process, without jets, is given by the  NeXSPheRIO model~\cite{Takahashi:2009na,Hama:2009vu,Andrade:2009em,Andrade:2010sd,Andrade:2010xy}. Simulations with a single flux tube on the periphery of the collision zone in the initial state with subsequent hydrodynamic evolutions produce two-particle correlation structures qualitatively consistent with the observations on both the near and away side. They further show that the ridge production is local--it is the local peripheral tube that generates the correlation structure, not the global structure of the initial conditions~\cite{Hama:2012ap}. The model seems to be able to describe the observed dependence of the ridge magnitude as a function of the trigger particle direction relative to the reaction plane~\cite{Qian:2012qn}. Using the one-tube model, the authors demonstrate that the reaction-plane dependent ridge in the model is due to an interplay between the background elliptic flow caused by the initial state global geometry and the effect resulted from the fluctuating flux tubes.

While the one-tube initial condition is illustrative, more realistic initial condition fluctuations are required for quantitative comparison to data. Hydrodynamic calculations use the Glauber model~\cite{Miller:2007ri}, 
the MC-KLN model~\cite{Drescher:2006ca}, or the IP-Glasma model~\cite{Schenke:2012wb} as input to describe the initial energy density and its fluctuations in heavy-ion collisions. The Glauber initial condition does not give a more specific physics description beyond the probabilistic nature of the binary nucleon-nucleon collision and particle production from it. The KLN or Glasma initial conditions describe the transverse energy density profile. Since the higher order harmonics are generated by the fluctuations in the initial state, the $v_n$ measurements themselves can be used to help distinguish between models of the initial state. A PHENIX study~\cite{Adare:2011tg} finds that a calculation based on the Glauber model is better able to describe the centrality dependence of $v_3$ than one based on the MC-KLN model.  The IP-Glasma model is able to successfully describe $v_n$ measurements at both RHIC and the LHC~\cite{Gale:2012rq} as shown in the left plane of Fig.~\ref{fig:phenix_v3}.

Of course the $\eta/s$ value, among others, is also an input to the hydrodynamic calculations and determining this quantity precisely from data is one of the main goals of these measurements. Viscosity is known to reduce the final-state flow~\cite{Romatschke:2007mq,Molnar:2008xj}. It is shown by event-by-event viscous hydrodynamic calculations that $n>$2 flow is more damped by viscosity than elliptic flow, thereby providing a more sensitive probe to $\eta/s$~\cite{Schenke:2010rr,Song:2010mg,Qiu:2011iv,Schenke:2011bn}.  
Comparison between elliptic flow data and viscous hydrodynamic calculations suggest that the $\eta/s$ cannot be more than a few times larger than the conjectured quantum lower limit of $1/4\pi$ in relativistic heavy-ion collisions at RHIC~\cite{Kovtun:2004de,Romatschke:2007mq,Song:2010mg,Schenke:2011bn,Luzum:2012wu}. 
The calculation shown in Fig.~\ref{fig:phenix_v3} by the IP-Glasma model~\cite{Gale:2012rq} requires different values of $\eta/s$ to fit the RHIC and LHC data, suggesting the sensitivity of $v_n$ measurements to differences between the matter created at RHIC and the LHC. 

It is important to point out that the extraction of $\eta/s$ has assumed negligible nonflow contributions in experimental measurements. The success of hydrodynamics with a particular $\eta/s$ to describe data does not necessarily mean there is no nonflow contributions in experimental data. 
The question how much of the measured anisotropy is due to hydrodynamic flow and how much is due to intrinsic nonflow correlations will, ultimately, have to be answered experimentally. 

In flow measurements using Eqs.~(\ref{eq:vn_fact}) and (\ref{eq:vn_fact_fluc}), flow factorization is explicitly assumed. Flow factorization is naively expected where single particles are correlated to a perceived common harmonic plane. It is pointed out that this perception may not be true; harmonic planes may decorrelate over $\deta$~\cite{Bozek:2010vz,Petersen:2011fp,Xiao:2012uw}. This will break factorization of flow harmonics over $\deta$, making the $\eta$-gap method of flow measurements less viable as aforementioned. 
It is also pointed out that particle correlations resulted from quantum fluctuations in the nuclear wavefunctions, with hydrodynamic flow only (without any nonflow contamination), can break factorization to some degree~\cite{Gardim:2012im}.
More recently, it is shown~\cite{Heinz:2013bua} by event-by-event hydrodynamic calculations that the event plane depends even on $\pt$ (at the same $\eta$) of particles used for reconstruction of the event plane. While the $\eta$-dependent event-plane may root in the $\eta$-variation of initial geometry configuration, the $\pt$-dependent event-plane may indicate dynamic, $\pt$-dependent responses and sensitivities to the initial geometry configuration. 

The $\eta$ and $\pt$ dependent harmonic planes, with pure hydrodynamics, break the factorization of flow harmonics:
\begin{eqnarray}
\Vn{\pt,\eta;\ptref,\etaref}
&=&\langle v_n(\pt,\eta)v_n(\ptref,\etaref)\rangle\langle\cos n[\psi_n(\pt,\eta)-\psi_n(\ptref,\etaref)]\rangle\nonumber\\
&\neq&\langle v_n(\pt,\eta)v_n(\ptref,\etaref)\rangle\,.\label{eq:vn_EPdecor}
\end{eqnarray}
This will make flow measurements by Eqs.~(\ref{eq:vn_fact}) and (\ref{eq:vn_fact_fluc}), which explicitly assume flow factorization, problematic.
Note that in Eq.~(\ref{eq:vn_EPdecor}) we have assumed that flow magnitude fluctuation and event-plane decorrelation are independent of each other, which seems to be supported by recent event-by-event hydrodynamic calculations~\cite{Heinz:2013bua}.
Given event-plane decorrelation, two-particle correlation background should be 
\begin{equation}
\langle v_n(\ptt,\etat)v_n(\pta,\etaa)\rangle\langle\cos n[\psi_n(\ptt,\etat)-\psi_n(\pta,\etaa)]\rangle\,.\label{eq:vnvn_bkgd}
\end{equation}
Experimentally the flow parameters are measured by correlating trigger and associated particles, separately, with reference particles. The flow background used in correlation subtraction is 
\begin{eqnarray}
&&\frac{\langle v_n(\ptt,\etat)v_n(\ptrefa,\etarefa)\rangle\langle v_n(\pta,\etaa)v_n(\ptrefb,\etarefb)\rangle}{\langle v_n(\ptrefa,\etarefa)v_n(\ptrefb,\etarefb)\rangle}\times\nonumber\\
&&\frac{\langle\cos n[\psi_n(\ptt,\etat)-\psi_n(\ptrefa,\etarefa)]\rangle\langle\cos n[\psi_n(\pta,\etaa)-\psi_n(\ptrefb,\etarefb)]\rangle}{\langle\cos n[\psi_n(\ptrefa,\etarefa)-\psi_n(\ptrefb,\etarefb)]\rangle}\,.\label{eq:vnvn_meas}
\end{eqnarray}
The inequality between the quantities of Eq.~(\ref{eq:vnvn_bkgd}) and Eq.~(\ref{eq:vnvn_meas}) may be less severe than that in Eq.~(\ref{eq:vn_EPdecor}). However, the quantitative degree of the inequality, hence the effect on flow background subtraction in correlation studies, should be further investigated. 

\section{Possible Effects of Jet-Medium Interactions}\label{sec:jet}
As discussed in the previous section, hydrodynamical higher order $v_n$ harmonics could qualitatively explain the structures seen in two-particle correlations, it is natural to ask if hydrodynamic $v_n$ can account for all of the structures seen in two-particle correlations at low and intermediate $p_T$.  Given that there are strong jet correlations in \pp\ collisions in the same $p_T$ region~\cite{Adler:2002tq,Adams:2005ph,Adler:2006sc} and there are unambiguous signals of jet-quenching in two-particle correlations at high $p_T$~\cite{Adler:2002tq,Adams:2006yt,Adare:2010ry}, jet induced effects must be present. In fact, the primary goal of dihadron and multi-hadron correlations with high $\pt$ trigger particles is to study the energy loss mechanisms due to interactions of energetic partons with the QGP medium. A high-$\pt$ trigger particle biases its parent parton to emit from the surface region, allowing maximal path-length of the away-side partner parton to traverse the medium, hence the maximal interaction strength.

\begin{figure}[hbt]
\begin{center}
\includegraphics[width=\textwidth]{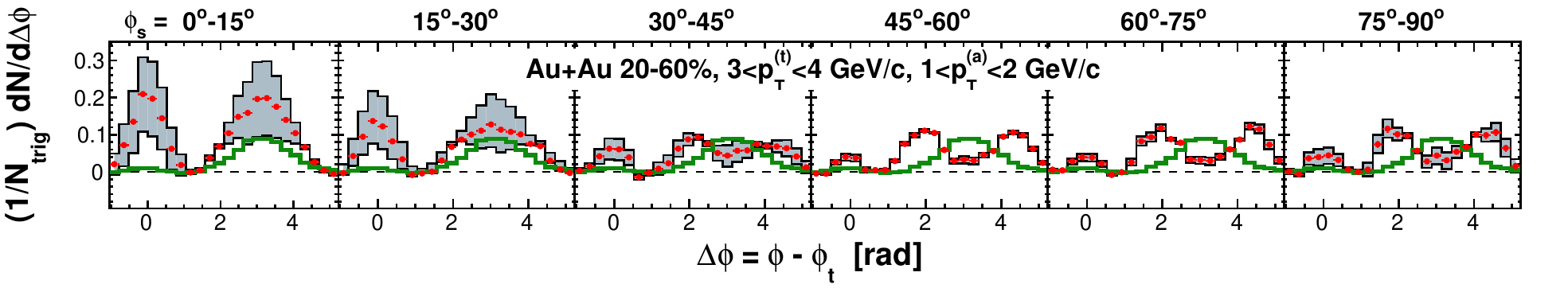}
\includegraphics[width=\textwidth]{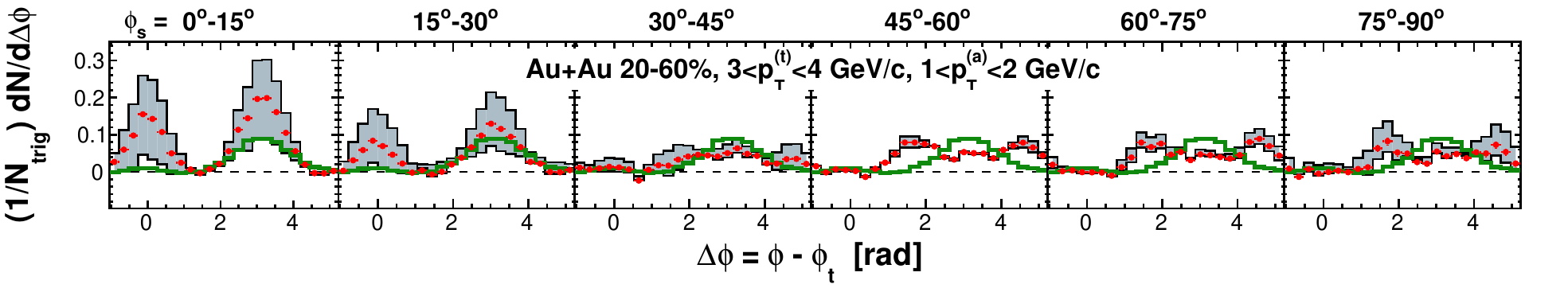}
\end{center}
\caption{Dihadron correlations at $|\Delta\eta|>0.7$ in 20-60\% Au+Au collisions at 200~GeV with trigger particles in six slices of azimuthal angle relative to the event plane, $\phi_s=|\phi_t-\psi_{EP}|$. The upper panels have the $v_2$ and $v_4$ background subtracted, and the lower panels have the additional $v_3$ background subtracted. The trigger and associated particle $\pt$ ranges are $3<\pt<4$~GeV/$c$ and $1<\pt<2$~GeV/$c$, respectively. 
Systematic uncertainties due to flow subtraction are shown in the shaded areas. 
The inclusive dihadron correlations from \dau\ collisions (thick histograms) are superimposed for comparison. From STAR~\cite{Agakishiev:2010ur,Wang:2013jro}.}
\label{fig:STAR_1010.0690}
\end{figure}

In non-central collisions, one may vary the path-length of the away-side parton by selecting the high-$\pt$ trigger particle azimuthal angle relative to the event plane, $\phi_s$. Both STAR and PHENIX have studied dihadron correlations as a function of $\phi_s$~\cite{Agakishiev:2010ur,PHENIX:2013kia,Wang:2013jro}. Figure~\ref{fig:STAR_1010.0690} from STAR shows the correlation results at large $\deta>0.7$ for mid-central 20-60\% Au+Au collisions~\cite{Agakishiev:2010ur}. The event plane is reconstructed from particles at least 0.5 unit away in pseudorapidity from the trigger particle and with $\pt<2$~\gev\ but excluding the associated particle $\pt$ bin. The top panels of Fig.~\ref{fig:STAR_1010.0690} shows these correlations where the background, including $v_2$ and $v_4$ (measured with respect to $\psi_2$), has been subtracted. The bottom panels of Fig.~\ref{fig:STAR_1010.0690} shows the results after subtraction of additional $v_3$. The effect of the $v_4$ uncorrelated to the second harmonic plane, $\psi_2$, is negligible for the 20-60\% centrality of Au+Au collisions~\cite{Agakishiev:2010ur}. The systematic uncertainty bands correspond to the conservative estimate of systematic uncertainties on $v_2$ by the two- and four-particle cumulant methods. The smaller uncertainties for out-of-plane triggers are due to the fact that the magnitude of the trigger particle $\langle\cos2(\phi_{trig}-\psi_2)\rangle$, itself being negative, is smaller when the trigger particle $v_2$ is larger. As a result, the flow modulation, $v_2^{assoc}\langle\cos2(\phi_{trig}-\psi_2)\rangle$, is less sensitive to the uncertainty in $v_2$. The subtracted $v_3$ is obtained from two-particle cumulant measurement with an $\eta$-gap of 0.7. This is considered to be the maximum possible $v_3$ background contribution because the trigger-associated correlation is also constrained to be at $|\deta|>0.7$. 

As seen from the upper panels of Fig.~\ref{fig:STAR_1010.0690}, the near-side ridge correlation magnitude decreases with $\phi_s$. This is in contrast to the jet-correlated yield at small $\dphi$ and small $\deta$, obtained from the difference in the $\dphi$ correlations between at small and large $\deta$, which is independent of $\phi_s$. This suggests that the ridge is more connected to the bulk medium than jet production. With additional subtraction of $v_3$ in the lower panels of Fig.~\ref{fig:STAR_1010.0690}, the ridge correlation is largely reduced, but a small finite ridge seems to remain for in-plane triggers beyond the maximum allowed $v_2$ and $v_3$ backgrounds. 

As shown in Fig.~\ref{fig:STAR_1010.0690}, the general feature of the away-side correlation does not seem to change after $v_3$ subtraction. The away-side correlation is single-peaked for in-plane triggers and becomes double-peaked for out-of-plane triggers. 
What could eliminate the double-peak? One way would be larger $v_n$. The remaining correlation structures are typically 10\% of the anisotropic flow modulation, and hence, a 10\% increase in $v_2$ beyond the two-particle maximum $v_2$ measurements could explain them. For example, Luzum~\cite{Luzum:2010sp} showed by Fourier decomposition of the correlation functions that the correlations can be completely explained by increasing $v_2$ by approximately 10\% beyond the upper $v_2$ systematics. 
Here the upper systematic bound is the two-particle $v_n$ with $\eta$-gap of 0.7, which are the maximum $v_n$ for dihadron correlations at $|\deta|>0.7$. 

However, the event plane, although reconstructed with particles 0.5 unit away from the trigger particle in $\eta$, can still be influenced by jet-correlated particles, especially by those on the away side where back-to-back jets are not much correlated in $\eta$. It is thus still possible that the novel features remnant in dihadron correlations relative to the event plane are due to biases in the event-plane reconstruction. STAR~\cite{Wang:2013jro,Agakishiev:2010ur} has investigated this question utilizing event plane reconstructed including the $|\deta|<0.5$ particles, thus including stronger event-plane bias. The dihadron correlation results are not significantly different from those in Fig.~\ref{fig:STAR_1010.0690}. This suggests that possible event-plane bias may be small for results shown in Fig.~\ref{fig:STAR_1010.0690} or that a $\eta$ gap significantly larger than 0.5 is required between the particles and the event plane determination.

The event-plane dependent dihadron correlations indicate that the correlations may not be entirely explained by hydrodynamic anisotropic flow; nonflow correlations should exist. As shown in Fig.~\ref{fig:STAR_1010.0690} the away-side double-peak structure seems robust against a wide range of flow subtraction. The near-side ridge magnitude decreases from in-plane to out-of-plane is not affected by triangular flow. 
The remaining structures  may arise from modifications to the jet structure due to jet-medium interactions.

There are many jet-related theoretical proposals for the production mechanisms of the ridge and double-peak correlations. These theoretical models were proposed before the realization that hydrodynamic triangular flow may be responsible for the majority of the observed ridge and double-peak correlations. Their relevance to the remaining correlation structures after removal of odd harmonics needs further quantitative evaluation. These models can broadly divided into two categories. One relies on microscopic interactions of jets with flowing medium partons. The other is collective medium excitation by Mach cone shock waves generated by energy deposition of jets in the medium. 

\subsection{Mach-cone shock waves}\label{sec:machcone}
The possibility of Mach cone shock waves in nucleus-nucleus interactions is not new~\cite{Baumgardt:1975qv,Hofmann:1976dy}. An energetic particle deposits energy in a medium over a short period of time, disturbing the local energy density and pressure, producing medium excitations. When the energetic particle travels faster than the speed of sound of the medium, these excitations are propagated within the limited region of a Mach cone~\cite{Stoecker:2004qu,CasalderreySolana:2004qm,Ruppert:2005uz,CasalderreySolana:2006sq,Renk:2007rv,Shuryak:2009cy}. Decays of these excitations would result in double-peaked distributions of correlated particles. 

A wide range of calculations of isolated energetic partons interacting with the QGP medium indicates the possibility of Mach cone shock waves. These calculations can be grouped into two general categories. One is hydrodynamic calculations with source terms describing parton energy loss with linear response approximation~\cite{Chaudhuri:2005vc,Betz:2008ka,Neufeld:2010tz}. The other utilizes AdS/CFT correspondence which can only calculate effects of heavy quark energy loss~\cite{Yarom:2007ap,Chesler:2007an,Gubser:2007ga,Gubser:2007zr,Noronha:2008un,Khlebnikov:2010yt}. 
Mach cone shock waves can be generated by both collisional and radiative energy loss mechanisms~\cite{Neufeld:2009ep}. It is found that local excitations fall sharply as a function of the energy of the fast parton, making Mach cone production less likely for large parton energies or trigger particle momenta~\cite{Neufeld:2010xi}. In addition to Mach cones, calculations~\cite{Betz:2008ka} suggest a diffusion wake in the direction of the propagating jet, which may be stronger than the Mach cone strength. 
Mach cones are predicted to appear only in strongly coupled plasma~\cite{Ruppert:2005uz}. An $\eta/s$ ratio larger than 0.2 prevents the development of well-defined shock waves on time scales typical for relativistic heavy-ion collisions~\cite{Bouras:2009nn}. However, it is shown~\cite{Neufeld:2008fi} that Mach cones may also arise from partonic energy loss in a weakly coupled QGP.

Although a generic prediction of partonic energy loss in an opaque low viscous medium, 
Mach cone is shown to be unlikely observable in final-state hadrons after hadronization by the Cooper-Frye mechanism~\cite{Noronha:2008un}. Even before hadronization, multiple sources of energy deposition present in a parton shower will make well-defined formation of Mach cones unlikely~\cite{Neufeld:2011yh}. In addition, the collective flow of the QGP itself will likely affect the Mach cone, making its observation difficult~\cite{Satarov:2005mv}.

Besides Mach cone shock waves, other jet-related physics mechanisms have been proposed which can mimic a Mach cone like signal. For example, an away-side double-peak structure may be produced by two partons from $2\rightarrow3$ processes~\cite{Ayala:2012bv}, each generating head-shocks depositing energy in the medium instead of one parton in a Mach-cone scenario. Similarly gluon splitting could produced double-peak structure~\cite{Polosa:2006hb}. Similar effects can be produced by Cerenkov gluon radiation~\cite{Koch:2005sx,Dremin:2005an}. 

\subsection{Effects of medium on jets}\label{sec:deflected}
The first theoretical proposal~\cite{Armesto:2004pt} for the observed broad $\deta$ correlations is longitudinal flow pushing jet fragments to more forward or backward pseudorapidities. This scenario would produce an asymmetric ridge, larger at more forward (backward) $\eta$ for a forward (backward) trigger particle. This particular feature of the model has not been experimentally observed. 

Event-by-event fluctuations lead to hot spots--regions with larger than average initial transverse parton density~\cite{Ma:2010dv,Qin:2011uw}. The hot spots have relatively more extended span in longitudinal than transverse direction and are shown by hydrodynamic evolution to generate a two-particle correlation peak that falls off more slowly in the pseudorapidity than the azimuthal direction~\cite{Springer:2012iz}. However, after subtracting the effects of hot spots, the authors of Ref.~\cite{Ma:2010dv} show with the AMPT (A Multi-Phase Transport) model~\cite{Zhang:1999bd} that the away-side correlation is still double-peaked. It is further shown that $\gamma$-hadron correlations, without the complications of harmonic flow background, exhibit double-peak away-side correlations~\cite{Li:2010ts,Ma:2010dv}. The double-peak structure may not be due to Mach cone emission, but jets deflected by medium flow event-by-event~\cite{Ma:2010dv}. Such conclusion is also reached by hydrodynamic study of jet-induced medium excitation with a realistic source term for energy-momentum deposition~\cite{Betz:2008ka,Betz:2010qh}. 

\begin{figure}[hbt]
\begin{minipage}{0.5\textwidth}
\begin{center}
\includegraphics[width=\textwidth]{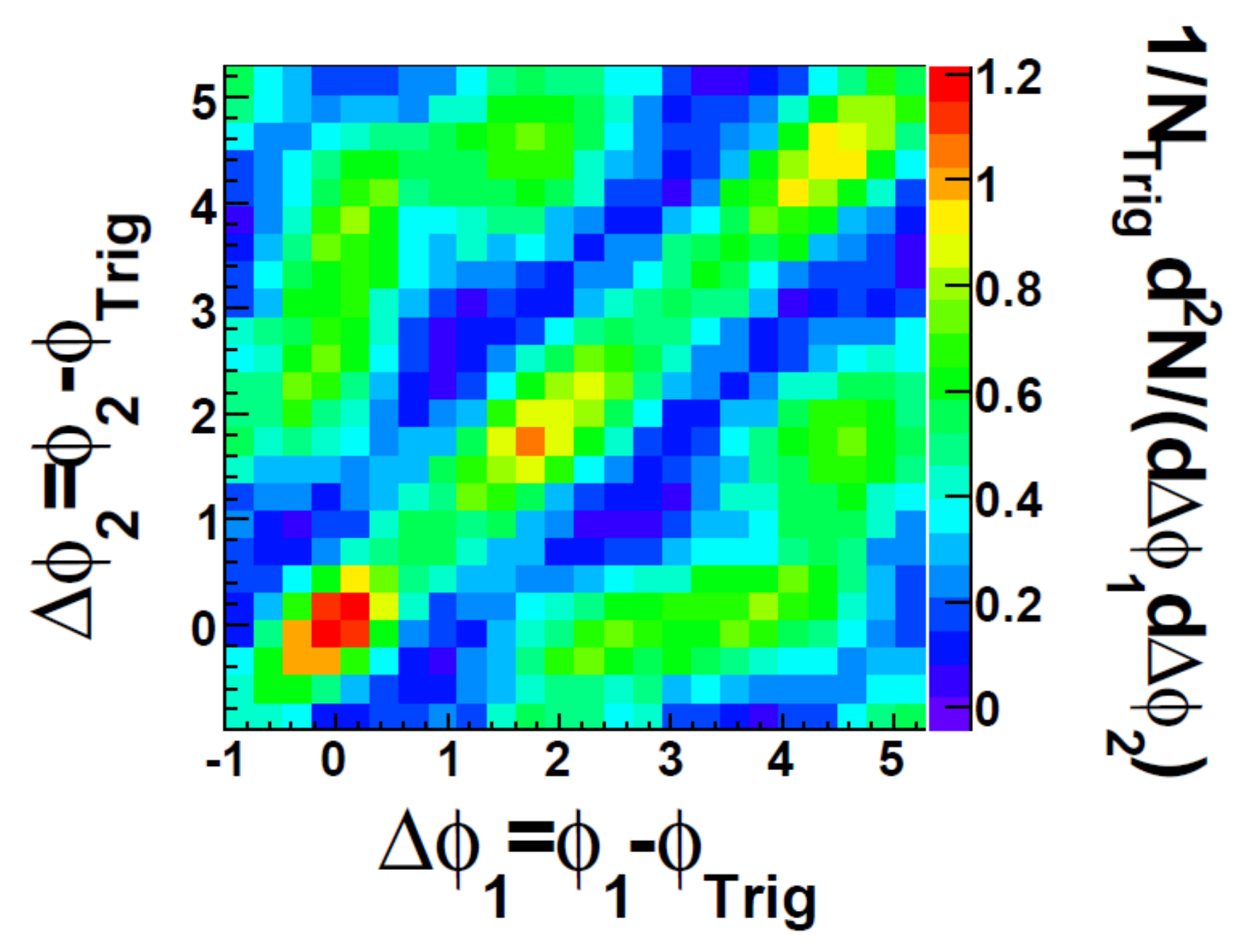}
\end{center}
\end{minipage}
\begin{minipage}{0.5\textwidth}
\caption{The $v_2$ and $v_4$ background subtracted three-particle correlations in 12\% central Au+Au collisions. From STAR~\cite{Abelev:2008ac}.}
\label{fig:STAR_PRL102_Fig2f}
\end{minipage}
\end{figure}

Evidence of deflected jets may be present in the three-particle correlation data~\cite{Abelev:2008ac,Ajitanand:2006is}. The three-particle correlations are initially measured with the potential to distinguish between the two leading scenarios proposed for the observed away-side double-peak in the two-particle correlations, namely deflected jets~\cite{Armesto:2004pt,Chiu:2006pu} and Mach-cone shock waves~\cite{Stoecker:2004qu,CasalderreySolana:2004qm,Ruppert:2005uz}.
In the deflected jet scenario, the away-side jet is collimated event-by-event but its direction is deflected by collective flow of the underlying event by varying angles. When summed over many events, the away-side correlations would appear broadened or double-peaked. In the Mach-cone scenario, the away-side particles are emitted preferentially along a cone about the away-side jet axis. On event-by-event basis the away-side correlations are double-peaked. 
Figure~\ref{fig:STAR_PRL102_Fig2f} shows the STAR three-particle correlation result in central Au+Au correlations~\cite{Abelev:2008ac}. Off-diagonal ($\dphi_1+\dphi_2=2\pi$) peaks are observed on the away side, corresponding to situations where the two correlated particles are emitted on the opposite sides of the back-to-back direction on an event-by-event basis. 
With the non-zero hydrodynamic odd harmonics, the off-diagonal three-particle correlation peaks are no longer unique signature of Mach-cone formation. Higher-order harmonics need to be subtracted in order to draw conclusions regarding Mach-cone formation. However, the away-side diagonal ($\dphi_1=\dphi_2$) peak strengths are stronger than the off-diagonal strengths in the measured three-particle correlations, as shown in Fig.~\ref{fig:STAR_PRL102_Fig2f}, whereas triangular flow can only produce equal-strength diagonal and off-diagonal peaks. This may be evidence of deflected jets on the away side of the trigger particle.

It has been argued that the ridge is produced by correlated emission between jet propagation and medium flow direction~\cite{Chiu:2008ht}. In this picture, the $v_2$ is due to surface emission of particles from semihard scatterings, and the ridge is produced by combining jet shower partons and medium partons that are aligned with each other. The model seems to have many desired features observed in data: the ridge is slightly harder than inclusive particles because a hard parton is involved, the ridge is strongest in the in-plane direction because the jet partons and medium partons are aligned, and it disappears in the out-of-plane direction because the partons are not aligned. The correlated emission model has a particular prediction: for trigger particles emitted at an angle from the reaction plane, because their momentum is not perfectly aligned with the flow direction which is perpendicular to the surface of the collision zone, the ridge correlation at large $\deta$ should be asymmetric about zero in $\dphi$. The asymmetries are opposite for trigger particles on the two sides of the reaction plane. This particular prediction seems confirmed by preliminary data from STAR~\cite{Konzer:2009xp}. However, further studies with respect to anisotropic flow, including high-order harmonics, are needed for a firm conclusion. Nevertheless, away-side asymmetric correlations are potentially important in probing jet-medium interactions because the path-length and the direction the away-side parton traverses the medium depend on the near-side trigger particle orientation relative to the reaction plane~\cite{Jia:2009tf}.

In the momentum kick model~\cite{Wong:2007pz,Wong:2008yh} 
the ridge is caused by medium partons which, after suffering a collision with jet, acquire a momentum kick along the jet direction and become correlated with the jet. The description requires the input of medium parton pseudorapidity density, and thus the comparison of this model to the ridge data can in principle yield information of the initial state pseudorapidity distribution of the medium partons. With the momentum kick model, there should be a ridge on the away side of the trigger particle as well, and the away-side ridge should be significantly stronger because of the larger probability of momentum kicks.


Unstable turbulent color field modes are shown to broaden jets more in $\eta$ than $\phi$~\cite{Majumder:2006wi,Mannarelli:2007gi,Dumitru:2007rp}. Quantitative calculations indicate, however, that the broadening in $\eta$ is insufficient to explain the experimentally observed approximately uniform ridge with high-$\pt$ particles. 

\section{Observation of the Ridge in Small Systems}
\begin{figure}[hbt]
\begin{center}
\includegraphics[width=0.45\textwidth,height=0.35\textwidth]{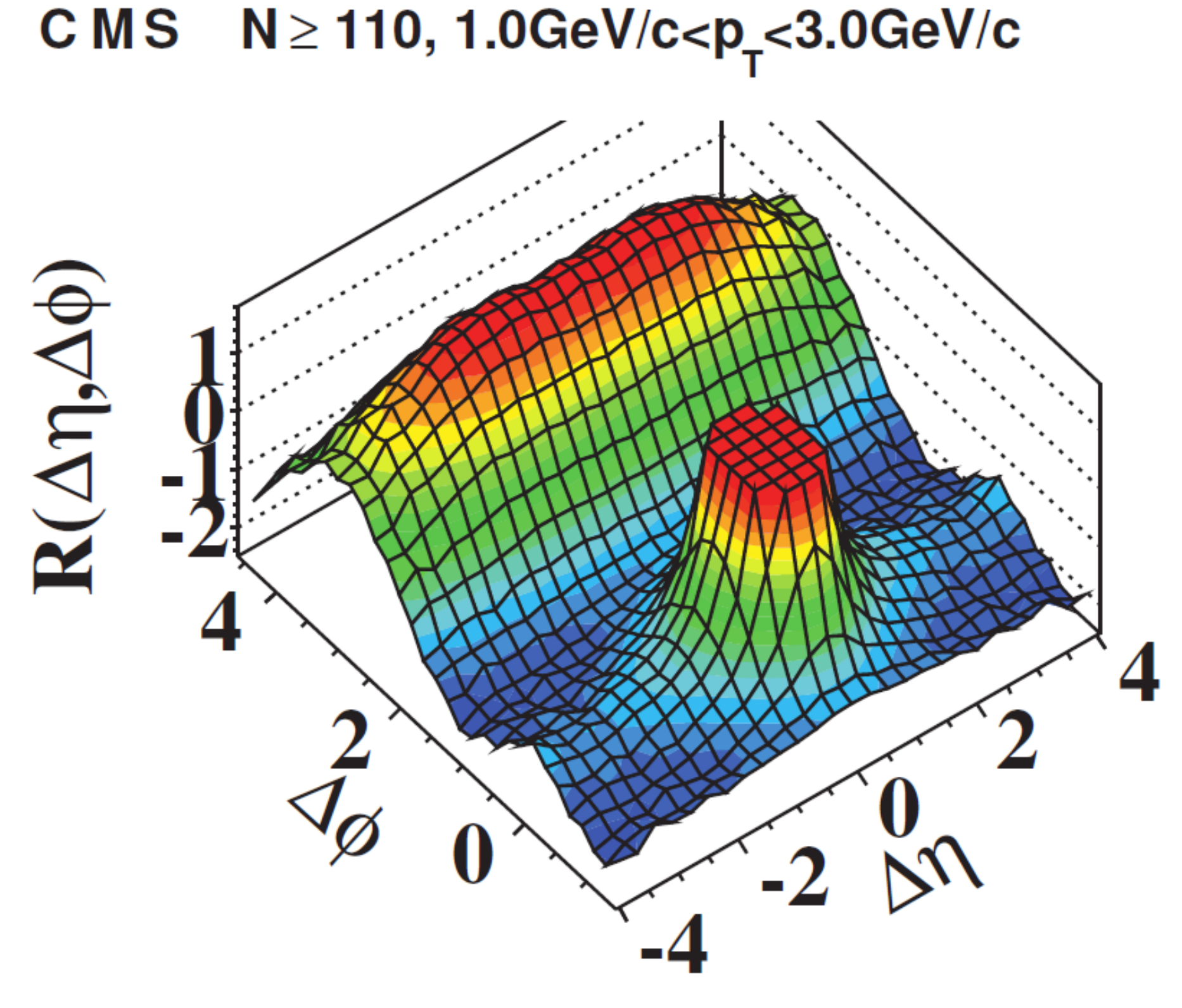}
\includegraphics[width=0.45\textwidth,height=0.35\textwidth]{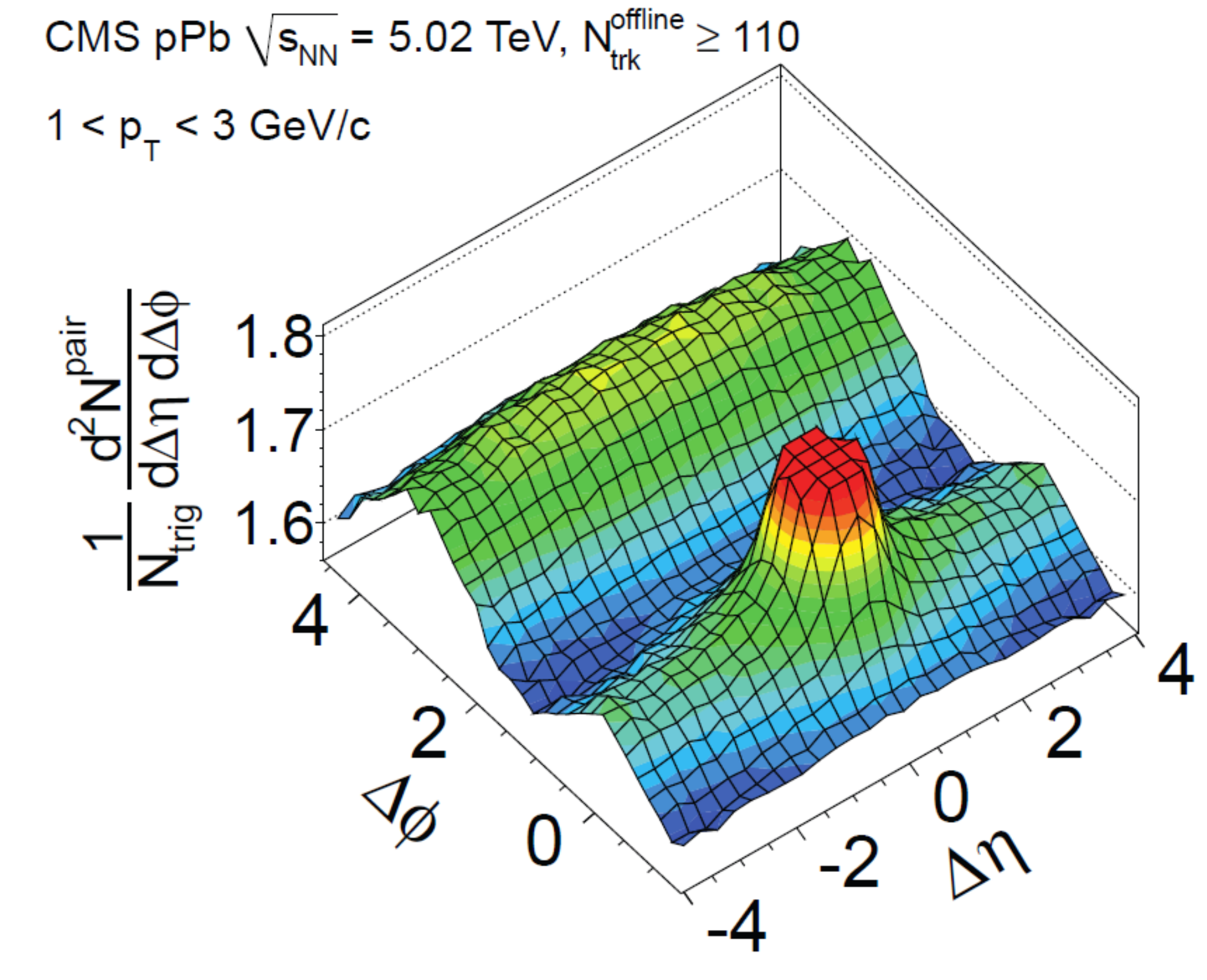}
\end{center}
\caption{Two-particle correlation functions in 7~TeV \pp~\cite{Khachatryan:2010gv} (left panel) and 5.02~TeV \ppb~\cite{CMS:2012qk} (right panel) collisions for pairs of charged particles with $1<\pt<3$~\gev. High-multiplicity events with $N^{\rm offline}_{\rm trk}>110$ (at $|\eta|<2.4$ and $\pt>0.4$~\gev) are used. Note a different correlation function definition is used for the left panel. 
From CMS~\cite{Khachatryan:2010gv,CMS:2012qk}.
}
\label{fig:CMS_pp_pPb}
\end{figure}

Unexpectedly, the CMS experiment at the LHC observed a long-range $\eta$ correlation in very high multiplicity \pp\ collisions at $\sqrt{s}=7$~TeV~\cite{Khachatryan:2010gv,Li:2012hc}. This is shown in the left panel of Fig.~\ref{fig:CMS_pp_pPb} where the trigger and associated particle $\pt$ ranges are both $1<\pt<3$~\gev. The event multiplicity cut is $N^{\rm offline}_{trk}>90$, with enough statistics owing to an online multiplicity trigger. With smaller $\pt$ or in events with smaller multiplicity, the ridge is not observed.  

\begin{figure}[hbt]
\begin{center}
\includegraphics[width=0.7\textwidth]{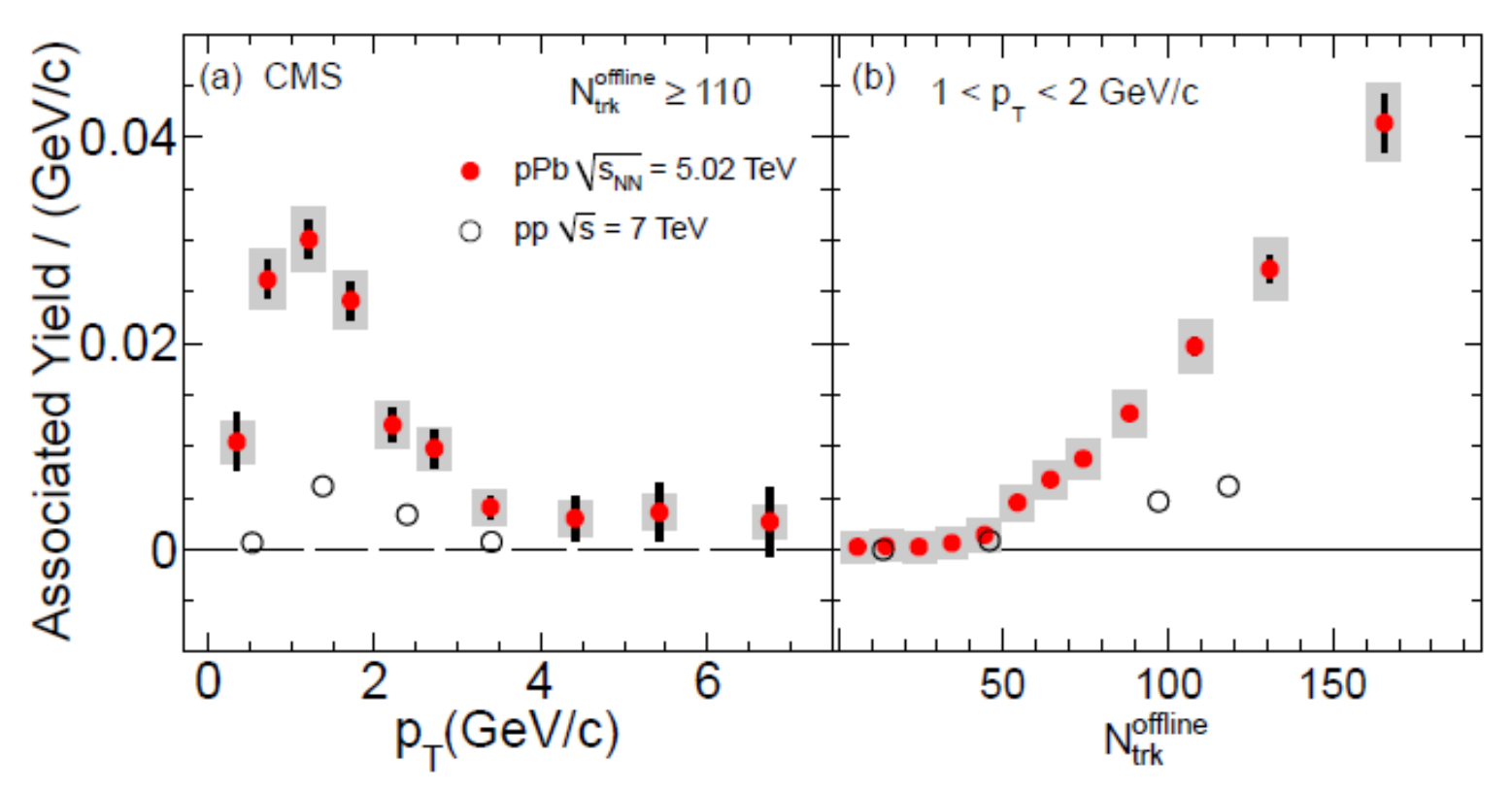}
\end{center}
\caption{The ridge yield (averaged over $2<|\Delta\eta|<4$ and integrated over the region $|\Delta\phi|<1.2$) in 7 TeV pp collisions (open circles) and 5.02 TeV \ppb\ collisions (solid circles). Panel (a) shows the $\pt$ dependence for events with $N^{\rm offline}_{\rm trk}>110$, and panel (b) shows the $N^{\rm offline}_{\rm trk}$ dependence for $1<\pt<2$~\gev. The $\pt$ selection applies to both particles. Error bars are statistical and shaded boxes are systematic uncertainties. From CMS~\cite{CMS:2012qk}.}
\label{fig:CMS_ridge}
\end{figure}

A ridge correlation in proton-lead (\ppb) collisions at $\snn=5.02$~TeV is observed by all three LHC experiments~\cite{CMS:2012qk,Abelev:2012ola,Aad:2012gla}. The result from CMS~\cite{CMS:2012qk} is depicted in the right panel of Fig.~\ref{fig:CMS_pp_pPb}. The ridge in \ppb\ is significantly larger than in \pp\ with the same multiplicity and $\pt$ cuts. 
The trends of the ridge magnitude as a function of $\pt$ and event multiplicity are, however, similar between \pp\ and \ppb\ collisions. This is shown in Fig.~\ref{fig:CMS_ridge}. 
The maximum associated yield is found at $\pt=1$-2~\gev. 
The rise and fall of the $\pt$ dependence are typical of hadron production. 

\begin{figure}[hbt]
\begin{center}
\includegraphics[width=0.32\textwidth]{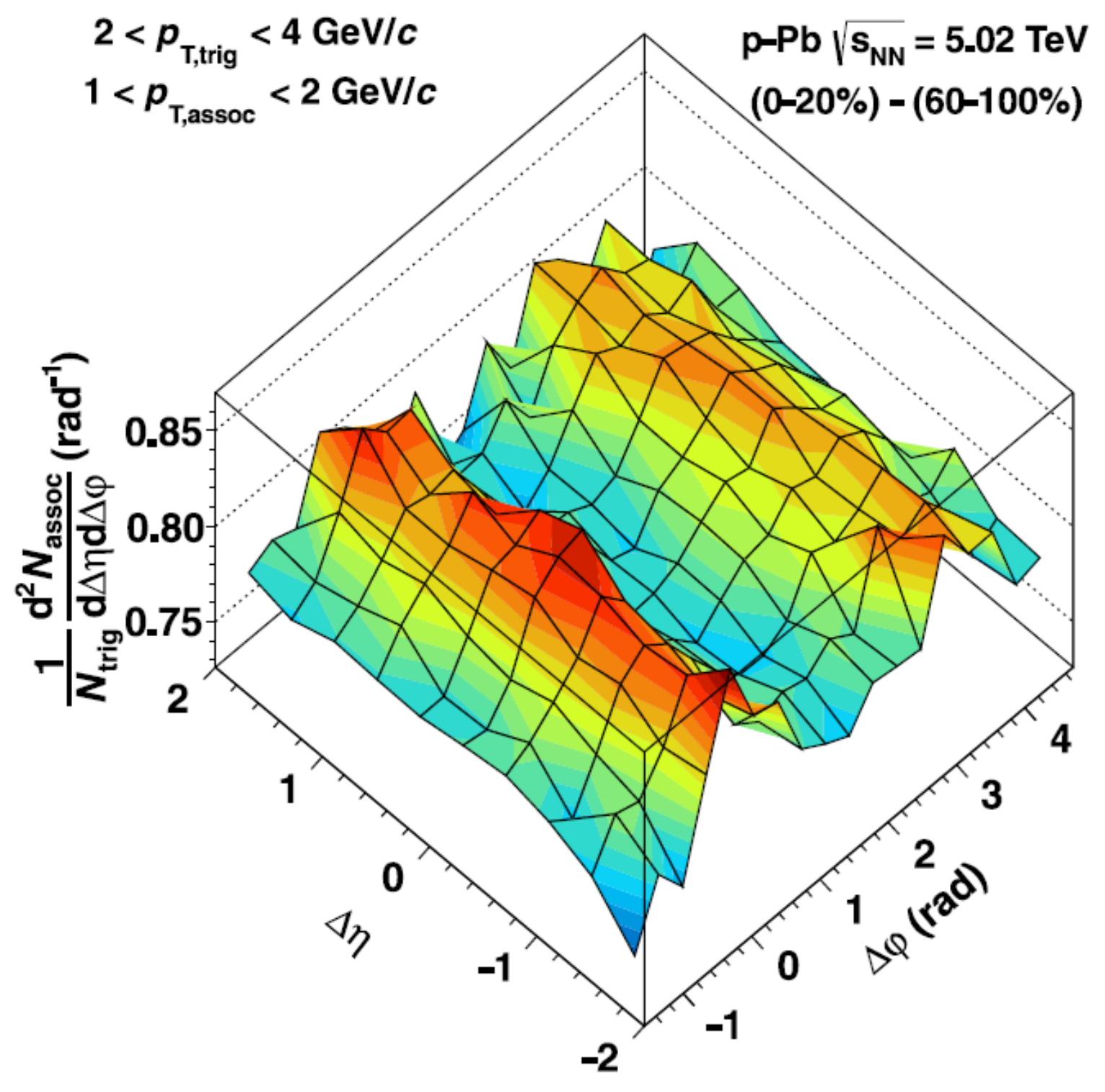}
\includegraphics[width=0.34\textwidth]{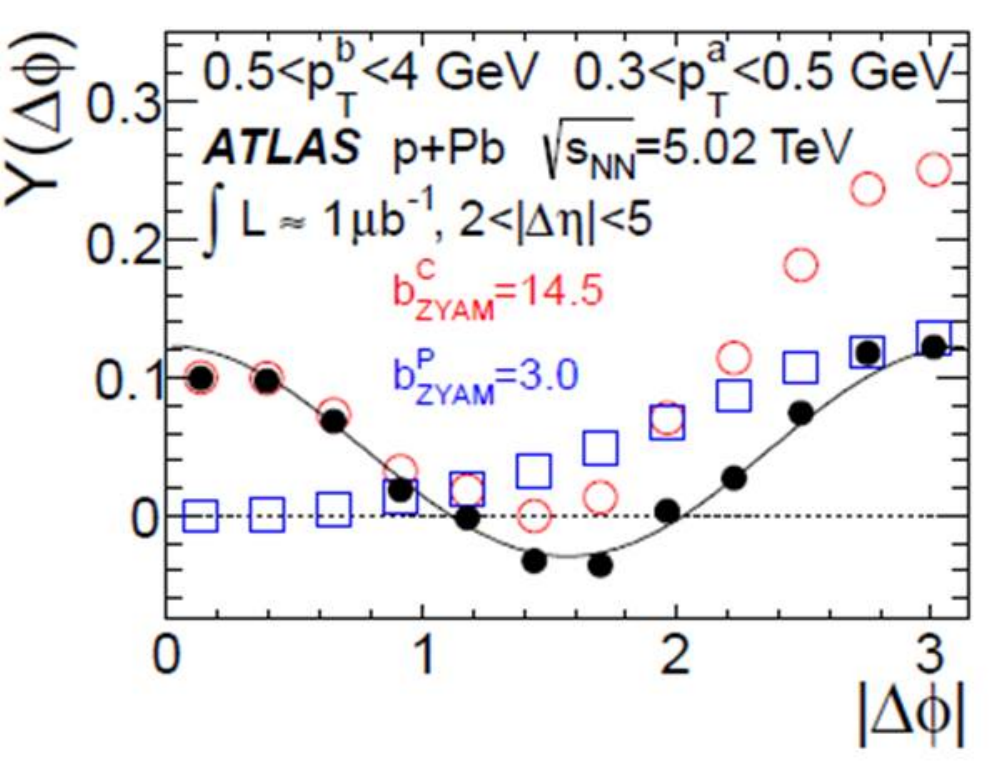}
\includegraphics[width=0.31\textwidth]{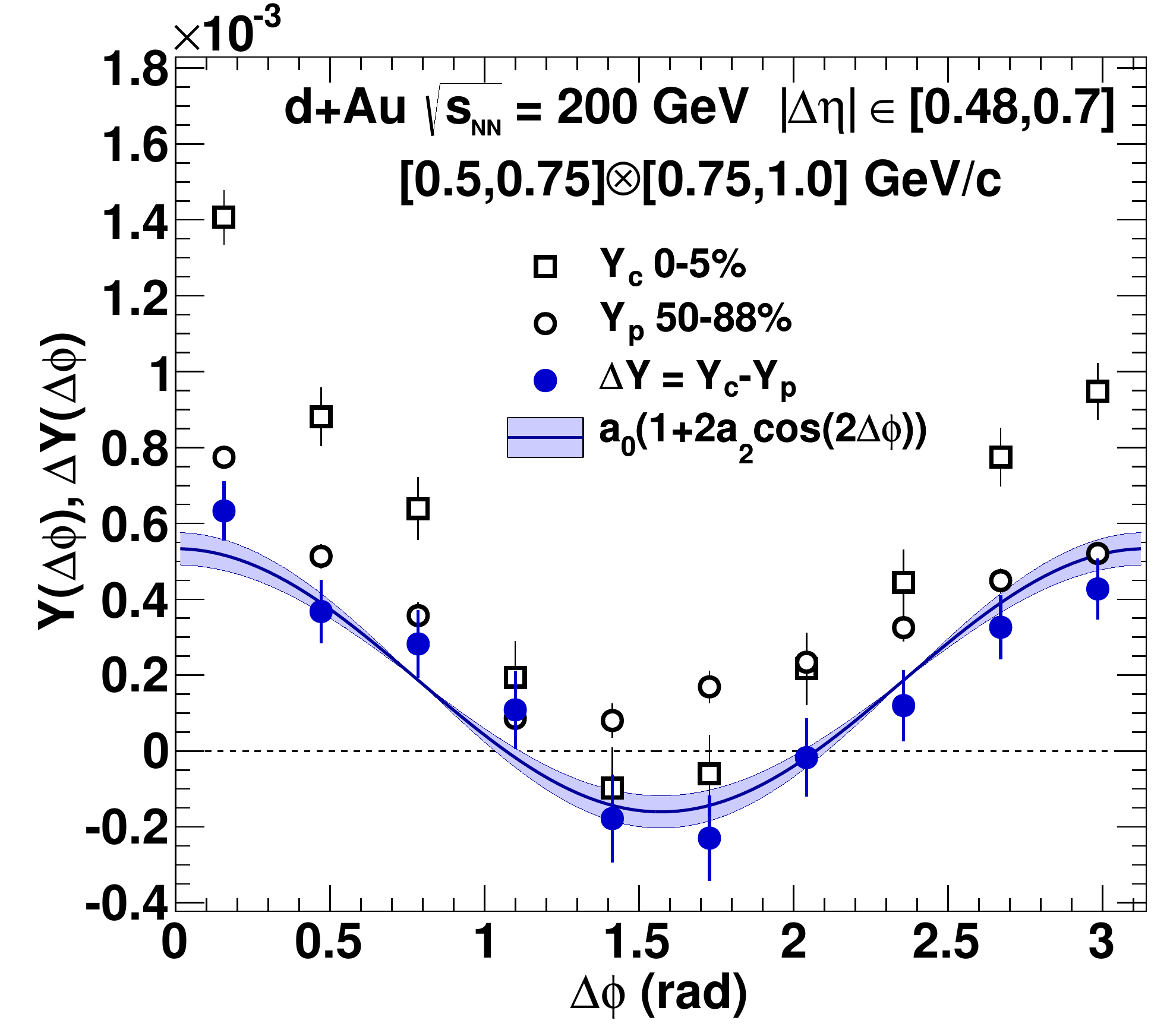}
\end{center}
\caption{(Left panel) ALICE result~\cite{Abelev:2012ola} on the difference in dihadron correlations between central 0-20\% and peripheral 60-100\% \ppb\ collisions. The trigger and associated $\pt$ ranges are $2<\ptt<4$~\gev\ and $1<\pta<2$~\gev, respectively. (Middle panel) ATLAS result~\cite{Aad:2012gla} of dihadron correlations in peripheral (48-100\%) and central (0-2\%) \ppb\ collisions and their difference; $0.5<\ptt<4$~\gev\ and $0.3<\pta<0.5$~\gev. (Right panel) PHENIX result~\cite{Adare:2013piz} of dihadron correlations in peripheral (50-88\%) and central (0-5\%) \dau\ collisions and their difference; $0.5<\ptt<0.75$ and $0.75<\pta<1$~\gev.}
\label{fig:ALICE_pPb}
\end{figure}

ALICE and ATLAS further investigated the excess, ridge correlation in high-multiplicity \ppb\ events by subtracting the dihadron correlation in low-multiplicity events to reduce jet contributions~\cite{Abelev:2012ola,Aad:2012gla}. It is found that the excess is approximately symmetric between $\Delta\phi\approx$~0 and $\Delta\phi\approx\pi$, and can be characterized largely by a $\cos2\dphi$ modulation (see Fig.~\ref{fig:ALICE_pPb}). This is surprisingly similar to heavy-ion collisions where the modulation is mainly attributed to collective elliptic flow. In the subtraction method, the jet-correlations from  peripheral collisions are subtracted from central collisions. Difference in the central and peripheral correlations (aside from the trivial multiplicity difference which has been removed by the ZYAM procedure) may be attributed to a new effect, whether a centrality dependent jet correlation or a new mechanism.  

\begin{figure}[hbt]
\begin{center}
\begin{minipage}{0.45\textwidth}
\includegraphics[width=\textwidth]{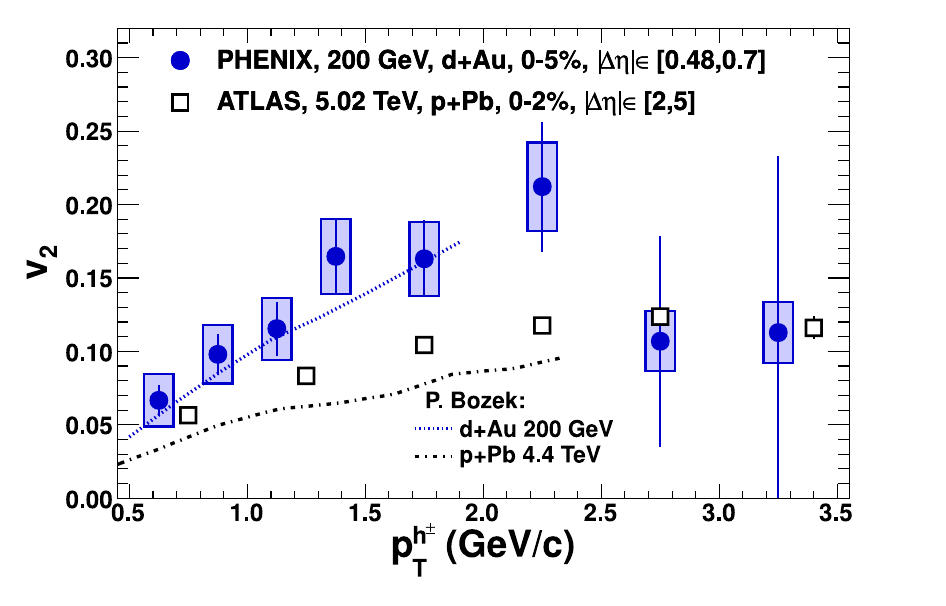}
\end{minipage}
\begin{minipage}{0.45\textwidth}
\caption{
The $v_2$ parameter extracted from 0-5\% central \dau\ collisions at $\sqrt{s_{NN}}$=~200~GeV by PHENIX~\cite{Adare:2013piz} and from 0-2\% \ppb\ collisions at 5.02~TeV by ATLAS~\cite{Aad:2012gla}.  The lines are calculations from a hydrodynamic model~\cite{Bozek:2011if}. 
From Ref.~\cite{Adare:2013piz}.
}
\label{fig:dAuv2}
\end{minipage}
\end{center}
\end{figure}

Motivated by the LHC \ppb\ data, the PHENIX experiment~\cite{Adare:2013piz} has analyzed the \dau\ data at $\snn=200$~GeV at RHIC using the same analysis procedure of subtracting peripheral data from central data, though without the large pseudorapidity coverage of the LHC experiments.
An example of the ZYAM background subtracted correlation functions in central and  peripheral collisions as well as their difference is shown in Fig.~\ref{fig:ALICE_pPb} (right panel). The second Fourier coefficient is directly calculated from the difference distribution and is found to be as large as 0.01, qualitatively similar to that observed in \ppb\ collisions at the LHC. The extracted $v_2$ values as a function of $p_T$ for the 5\% most central events are shown in Fig.~\ref{fig:dAuv2} along with the ATLAS values~\cite{Aad:2012gla} for the 2\% most central \ppb\ collisions.  The PHENIX $v_2$ is significantly larger and agrees well with a hydrodynamic calculation corresponding to the same centrality selection~\cite{Bozek:2011if}. 

\subsection{Possibility of collective flow in small systems}
It is important to note that the ridge correlation observed in \pp\ and \ppb\ systems is after the subtraction of a uniform background in $\dphi$, unlike in heavy-ion collisions where the elliptic flow background is subtracted. 
Collective flow is not usually associated with elementary \pp, small system \ppb\ or \dau\ collisions. Hydrodynamics imply small mean free path and large interaction cross sections. It is applicable only to systems considerably larger than the mean free path of the constituents, and thus is generally not expected to work in very peripheral heavy-ion or small system collisions. However, the ridge results are suggestive of hydrodynamic effects in  \pp\ and $p(d)$+A collisions. A number of authors have described the \pp\ ridge with hydrodynamic calculations~\cite{Bozek:2010pb,Werner:2010ss}. 

\begin{figure}[hbt]
\begin{center}
\includegraphics[width=0.4\textwidth]{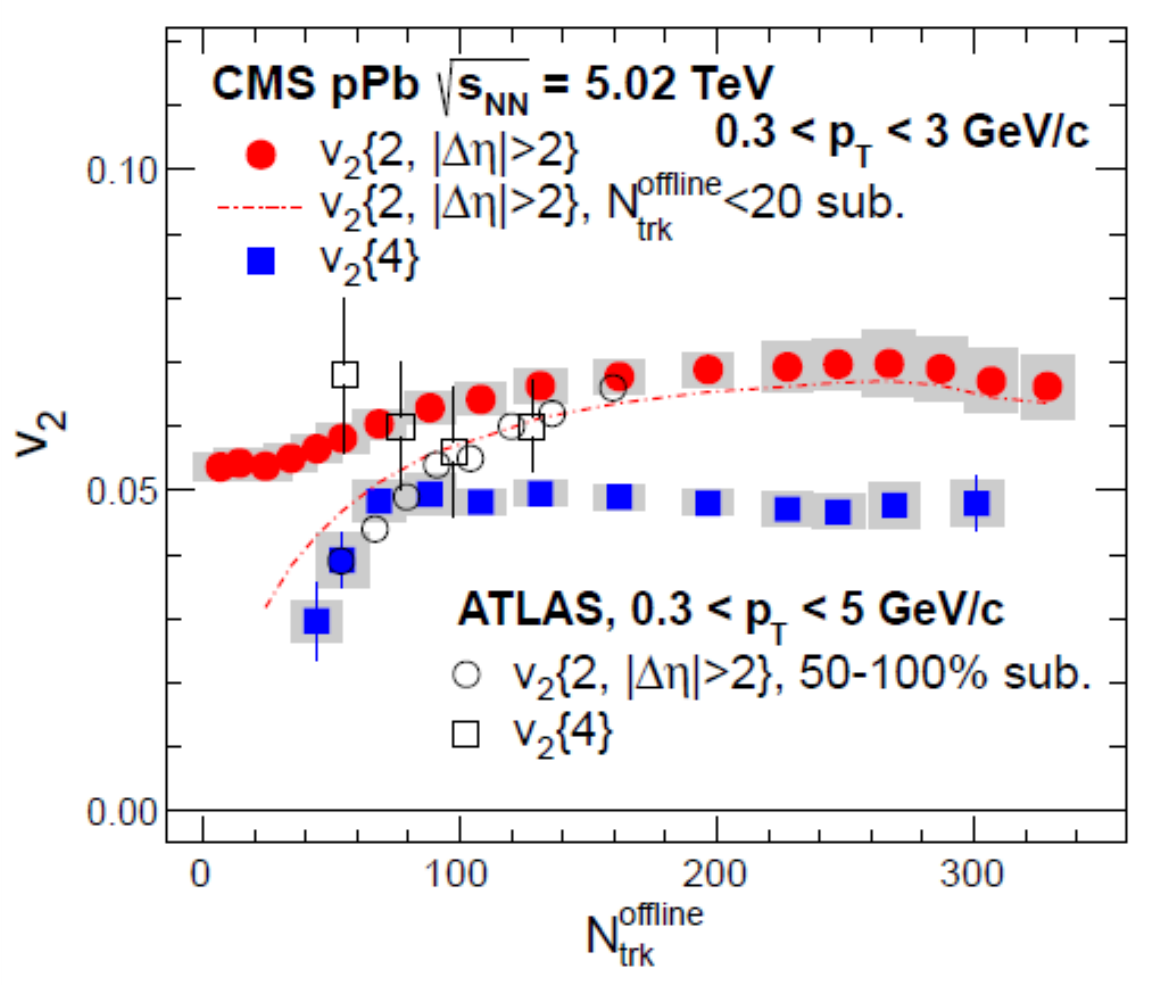}
\includegraphics[width=0.39\textwidth,height=0.335\textwidth]{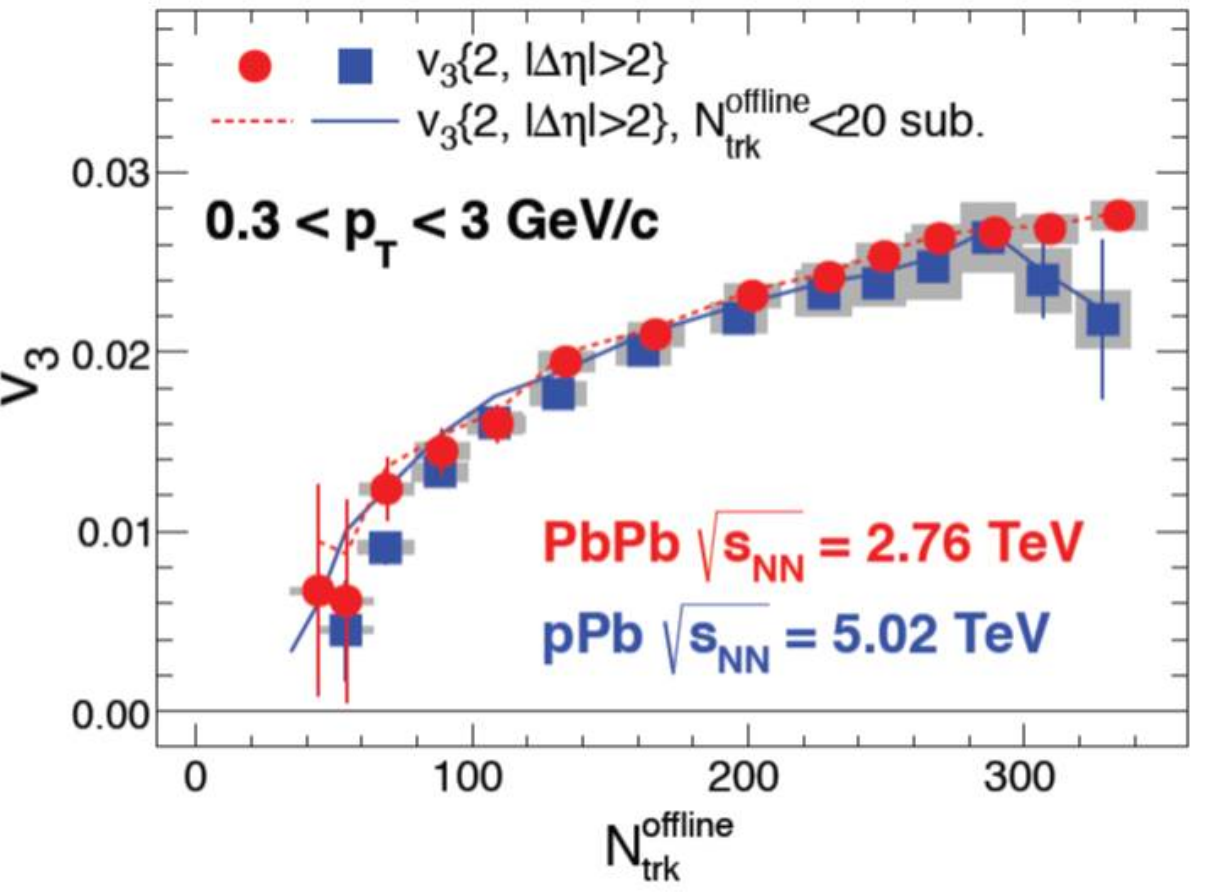}
\end{center}
\caption{(Left panel) The two- and four-particle cumulant measurements of $v_2$ in \ppb\ collisions by ATLAS~\cite{Aad:2013fja} and CMS~\cite{Chatrchyan:2013nka}. (Right panel) The two-particle cumulant measurement of $v_3$ in \ppb\ and Pb+Pb collisions as a function of event multiplicity by CMS~\cite{Chatrchyan:2013nka}. From Ref.~\cite{Chatrchyan:2013nka}.}
\label{fig:CMS_pPb_vn}
\end{figure}

The LHC experiments have systematically studied the Fourier coefficients of the two-particle ridge correlations. These Fourier coefficients would be the anisotropic flow squared in the absence of nonflow contributions, such as jets. In order to reduce nonflow contribution, four-particle cumulant method is employed to analyze \ppb\ data. Figure~\ref{fig:CMS_pPb_vn} left panel shows the two- and four-particle cumulant $v_2$ results as a function of event multiplicity. Significant four-particle $v_2$ is observed which is customarily attributed to collective anisotropic flow. With the relatively low multiplicity of \ppb\ collisions, the extent of nonflow contributions to the four-particle $v_2$ remains an open question. However, the four-particle $v_2$ is independent of event multiplicity except in very peripheral collisions, inconsistent with nonflow contributions. On the other hand, one would naively expect a larger flow in higher multiplicity events due to stronger hydrodynamic effects. The right panel of Fig.~\ref{fig:CMS_pPb_vn}  shows a comparison of the $v_3$ obtained from the third Fourier harmonic coefficients of the ridge correlation functions in \ppb\ and Pb+Pb collisions at similar multiplicity. The similarity is astonishing as $v_3$ is widely considered a manifestation of initial geometry fluctuations in Pb+Pb collisions and very different geometries for equal multiplicity \ppb\ and Pb+Pb collisions are naively expected. 

Further investigations at both RHIC and the LHC will be required to understand if hydrodynamics are the source of the long range ridge correlations observed in small systems.
If the matter produced in these systems can be described by hydrodynamic models, it is natural to expect very peripheral heavy-ion collisions to be described by hydrodynamics as well.  If the possibility of a hydrodynamic interpretation of these data holds up, then understanding the limits of hydrodynamics in peripheral heavy-ion collisions will also be of great interest.  Also it will be interesting to investigate if the matter in $p(d)$+A is large enough to observe some of the effects of jet quenching seen in central heavy-ion data.

\subsection{The Color Glass Condensate\label{sec:cgc}}
While the $\cos\left(2\Delta\phi\right)$ shape shown in Fig.~\ref{fig:ALICE_pPb} is naturally evocative of hydrodynamics, it is not the only theoretical attempt to describe the data. Another possible explanation is the Color Glass Condensate (CGC)~\cite{McLerran:1993ka}.
%
The gluon density in nucleons grows rapidly with decreasing fraction of the nucleon's momentum, $x$, carried by the gluon. The gluon density cannot, however, infinitely grow due to unitarity and self-interactions of gluons. The gluons combine resulting in a saturation of the gluon density at small $x$ below a certain momentum, $Q_s$, called the saturation scale~\cite{McLerran:1993ni,McLerran:1993ka,Iancu:2001md,Iancu:2002xk,Iancu:2003xm,Gelis:2007kn,Gelis:2010nm}.
The saturation scale can be determined where the rate of one gluon splitting into two is comparable to the rate of two gluons recombining into one.
Because of the Lorentz contraction, the saturation scale increases in large nuclei by a factor of $A^{1/3}$ where $A$ is the atomic number of the nucleus.  This makes nuclear collisions, especially $(d)p$+A, attractive systems in which to look for CGC effects.

The CGC is an effective field theory which allows first principle calculation of $n$-point gluon correlation functions order-by-order in perturbation theory from QCD. 
The CGC framework has been successful in describing some of the experimental observations. The multiplicity density is well described~\cite{Kharzeev:2002ei}. The semi-hard processes are predicted to be suppressed~\cite{Kharzeev:2002pc}.
Moreover, the dihadron correlations between forward- and mid-rapidity~\cite{Braidot:2010ig,Adare:2011sc} and between forward and forward rapidities~\cite{Adare:2011sc,Li:2012bn} are strongly suppressed on the away side in central \dau\ relative to \pp\ collisions. These effects can be described by CGC calculations~\cite{Lappi:2012bq}.

\begin{figure}[hbt]
\begin{center}
\begin{minipage}{0.3\textwidth}
\includegraphics[width=\textwidth]{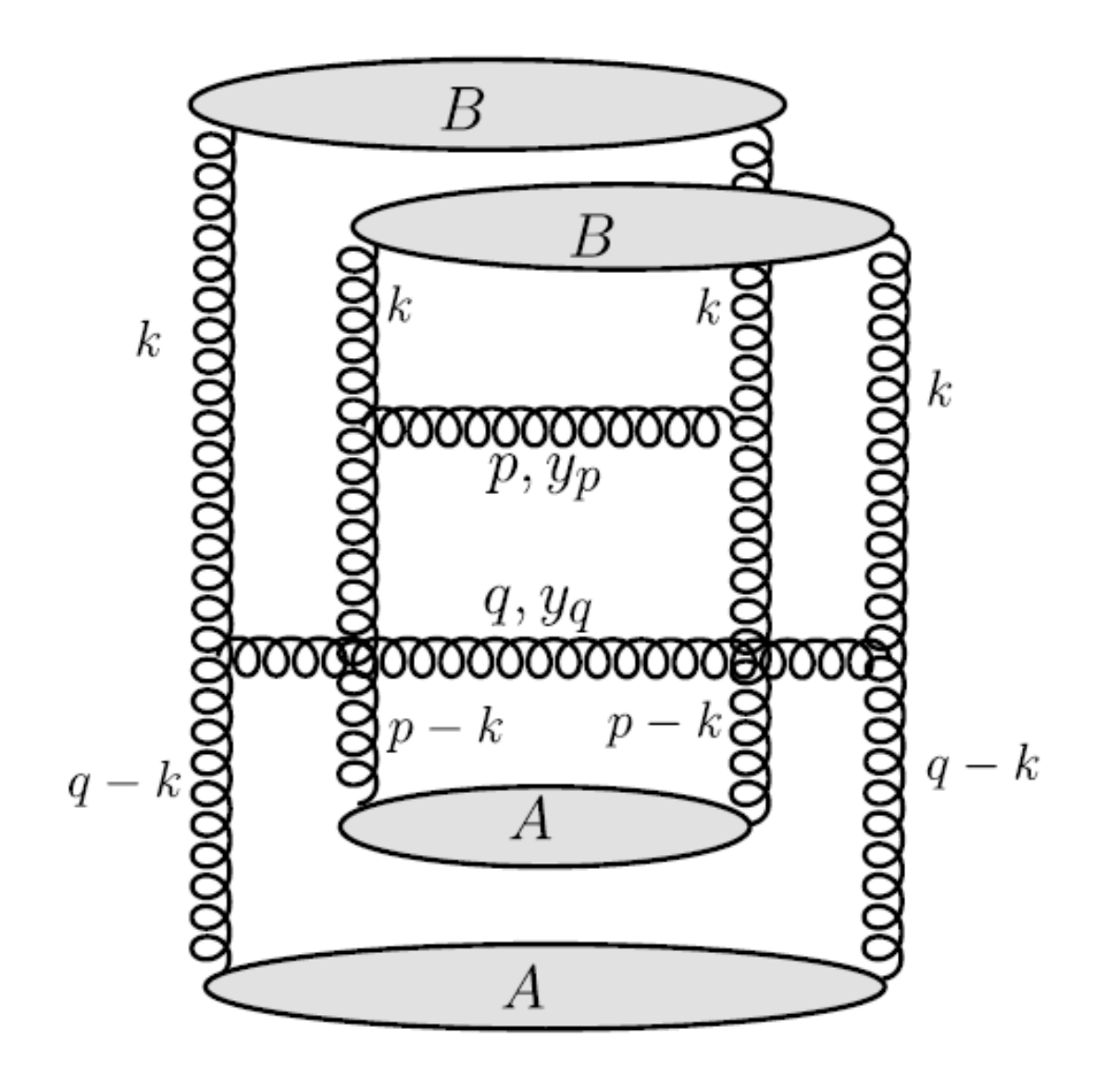}
\end{minipage}
\begin{minipage}{0.4\textwidth}
\caption{A typical two-particle production diagram in the CGC framework which gives an azimuthal angular collimation. From Ref.~\cite{Dumitru:2010iy}.}
\label{fig:Dumitru_PLB697}
\end{minipage}
\end{center}
\end{figure}

There is an intrinsic correlation in azimuthal angle coming from the two-particle production process in CGC, such as the one shown in Fig.~\ref{fig:Dumitru_PLB697}~\cite{Dumitru:2010iy}. There is only a single loop momentum $\mathbold{k_T}$ in this two-particle production process. Because the single gluon distribution peaks at the saturation scale $Q_s$, large probability is found for production of two particles with their momenta $\mathbold{p_T}$ and $\mathbold{q_T}$ parallel to each other such that $|\mathbold{p_T}-\mathbold{k_T}|\sim Q_s$ and $|\mathbold{q_T}-\mathbold{k_T}|\sim Q_s$, causing small angle correlations at $\dphi=0$. Because the correlations originate from the very early  times of the collision, $\tau_{\rm init.}$, they can persistent to large rapidity differences, $\Delta y=2\ln(\tau_{\rm f.o.}/\tau_{\rm init.})$ where $\tau_{\rm f.o.}$ is the particle freeze-out proper time.

It has been shown~\cite{Dumitru:2010iy,Dusling:2012iga,Kovchegov:2012nd} that the Glasma effective field theory can reproduce the \pp\ ridge and back-to-back correlations using the CGC two-gluon production process and the Balitsky-Fadin-Kuraev-Lipatov (BFKL) 
dynamics. The ridge structure emerges by going from the BFKL to the saturation region~\cite{Levin:2011fb}. The correlated two-gluon processes are only appreciable when the transverse distance, $1/Q_s$, is much smaller than the proton size~\cite{Dumitru:2010iy}.
Small impact parameter \pp\ collisions are thought to dominate high multiplicity events, which is why the ridge is observable only in high multiplicity events in the CGC framework. Because $Q_s$ gives the underlying event multiplicity and the magnitude of the ridge in \pp, the ridge measurements are potentially powerful to test CGC. 

The same framework can also reproduce the ridge in \ppb\ collisions~\cite{Dusling:2012cg,Dusling:2012wy,Dusling:2013oia} by varying the saturation scales in proton and lead where the latter depends on the number of participant nucleons from the lead target. High multiplicity events are described by large $Q_s$ in the proton (smaller impact parameter) and/or large number of participant nucleons ($N^{\rm Pb}_{\rm part}$) from the lead. The high multiplicity requirement is less efficient to increase $Q_s$ in the proton (i.e. triggering on small impact parameter) in \ppb\ than in \pp\ because it is highly improbable to have the nucleon-nucleon encounters in a \ppb\ collisions to be all at small impact parameters. The underlying event multiplicity and the ridge are then the result of nontrivial interplay between proton $Q_s$ and lead $N^{\rm Pb}_{\rm part}$. Thus the \ppb\ ridge data are a less stringent test of the CGC because there are two unknowns, $Q_s$ and $N^{\rm Pb}_{\rm part}$, the combination of which describes the ridge data as a function of multiplicity. However, if the ridge in \ppb\ is a result of CGC, the data give an effective measure of the $Q_s$ and $N^{\rm Pb}_{\rm part}$.

The multi-gluon correlations in the initial state~\cite{Gelis:2008sz} from CGC are insufficient to yield the large ridge correlations observed in central heavy-ion collisions. It is suggested that the ridge in heavy-ion collisions is a net effect of the initial state multi-gluon correlations and the final-state collective radial flow. The radial flow boosts the correlated particles from initial state into a small opening angle in $\phi$~\cite{Dumitru:2008wn,Dusling:2009ni,Gavin:2008ev,Moschelli:2009tg}. Similar spirit explanation of the ridge is suggested in Ref.~\cite{Voloshin:2003ud,Shuryak:2007fu} where the particles produced in an underlying \pp\ collision, which are correlated, are boosted by radial flow in heavy-ion collisions into a narrow opening in $\phi$ angle. As discussed in Sec.~\ref{sec:vn} the ridge correlations in heavy-ion collisions may be dominated by collective flow background, and are thus not as a decisive test of CGC as originally anticipated.
\section{Future Prospects}
Two-particle correlations in heavy-ion collisions provide a powerful tool to study the QGP medium properties. It is likely that both hydrodynamic flow and jet-medium interactions are present. They provide a unique opportunity to measure both the bulk properties of the system and the medium induced jet modifications at moderate $p_T$. In order to study jet-medium interactions, a careful subtraction of hydrodynamic flow should be performed. Likewise, in order to extract transport properties of the QGP from comparisons of flow correlations to hydrodynamics, a careful study of nonflow contributions is needed. Experiments are well on their way to carry out both of these tasks to achieve next-order precision in our understanding of the QGP medium.


Effects of nonflow correlations need further investigations on both the theoretical and experimental side. Nonflow correlations can be effectively reduced by $\eta$-gap (or even eliminated if it is possible to use an $\eta$-gap wider than the inter-jet $\eta$ differences). However, potential $\eta$-gap (and $\pt$) dependent flow fluctuations should be examined. Measurements of $v_n$ from event-plane and two-particle cumulant methods as well as the $\eta$ and $\pt$ dependence of such measurements are of prime interest.

Dihadron correlations with respect to the reaction plane are also a powerful tool to study jet-medium interactions with the additional control parameter of the away-side path-length. There remain questions as to the effects of nonflow correlations between the trigger particle and the reconstructed event plane have on the final dihadron results. To reduce the potential bias it is imperative to make this measurement with trigger and event-plane particles well separated in pseudorapidity.

The dihadron correlation analyses should be repeated by respective experiments with thorough subtraction of all non-negligible $v_n$, with proper systematic uncertainties to reflect unknown nonflow contributions in the $v_n$ measurements. This seems a high priority task if any jet-medium interaction effects are to be learned from particle correlation measurements. 


Recent measurements of the ridge and $v_n$ phenomena in small systems are especially interesting.  
The two leading candidate mechanisms for the near-side ridge in small systems are the CGC and hydrodynamic flow. These two distinct physics scenarios likely yield different dependences of the ridge as a function of collision energy, making the comparison of RHIC and LHC data, with a factor of 25 difference in collision energy, of great importance. The \dau\ collision data at RHIC together with the \pp\ and \ppb\ data at LHC and future $p$+A data at RHIC will help distinguish between the two proposed physics mechanisms. 


\section{Conclusions}
The matter created in relativistic heavy-ion collisions is a complicated system. The phenomena highlighted in this review show effects that were unexpected prior to heavy-ion measurements at RHIC and the LHC. They are discovered as one tries to gain detailed understanding of particle production and correlations in relativistic heavy-ion collisions. The hydrodynamic behavior of the bulk matter underlies, and may be responsible for some of, the phenomena discussed here.  In order to gain further understanding of QCD parton-parton interactions, improvements in quantitative hydrodynamic calculations and measurements are of great importance.

Observation of the phenomena reviewed in this article relies on high quality data, clever experimental techniques, and the emerging theoretical understanding of the QGP. The most valuable handles for understanding these phenomena are the large collision energy difference between RHIC and the LHC and the wide range of collision species.  Throughout this review comparisons have been made between RHIC and the LHC as well as between large and small systems.  Increasingly, data from the RHIC beam energy scan are also adding important new insights and constraints.  

Lastly, the field is exploiting the invaluable advantage of two complementary facilities. The experiences learned from RHIC are applied to the LHC. The new insights learned from the LHC are advancing our understanding at RHIC. Data from both facilities are needed to constrain measurements of the temperature dependence of the properties of the QGP. The coexistence of the RHIC and LHC heavy-ion programs will prove essential to our quest of understanding of high temperature QCD.

\section{Acknowledgments}
I thank my colleagues for discussions and collaboration. I thank Anne Sickles for help in the early preparation of this review. This work is supported by US Department of Energy Grant No. DE-FG02-88ER40412.  
\bibliographystyle{iopart-num}
\bibliography{corr_review}
\end{document}